\documentclass[iicol]{sn-jnl}
\normalbaroutside %
\usepackage{etoolbox}
\makeatletter
\patchcmd{\ps@headings}
{\hbox to \hsize{\hfill Springer Nature 2021 \LaTeX\ template\hfill}}
{\hbox to \hsize{\hfill Preprint. Under review.\hfill}}
{}
{}
\patchcmd{\ps@headings}
{\hbox to \hsize{\hfill Springer Nature 2021 \LaTeX\ template\hfill}}
{\hbox to \hsize{\hfill Preprint. Under review.\hfill}}
{}
{}
\patchcmd{\ps@titlepage}
{\hbox to \hsize{\hfill Springer Nature 2021 \LaTeX\ template\hfill}}
{\hbox to \hsize{\hfill Preprint. Under review.\hfill}}
{}
{}
\makeatother

\usepackage{silence}
\usepackage{appendix}

\usepackage[main=english]{babel}
\usepackage[utf8]{inputenc} %
\usepackage[T1]{fontenc}    %
\usepackage{microtype}      %
\usepackage{xspace}
\usepackage{hyphenat} %
\usepackage{siunitx} %
\ifdefined\qty\else
  \ifdefined\NewCommandCopy
    \NewCommandCopy\qty\SI
  \else
    \NewDocumentCommand\qty{O{}mm}{\SI[#1]{#2}{#3}}
  \fi
\fi
\ifdefined\unit\else
  \ifdefined\NewCommandCopy
    \NewCommandCopy\unit\si
  \else
    \NewDocumentCommand\unit{O{}m}{\si[#1]{#2}}
  \fi
\fi

 \DeclareSIUnit\molec{molec}

\usepackage{chemarr} %
\usepackage{amssymb}
\usepackage{amsmath}
\usepackage{amsfonts}       %
\usepackage{amstext} %
\usepackage{mathtools}
\usepackage{nicefrac}       %
\usepackage{amsthm}         %
\usepackage{bbm}
\usepackage{empheq}
\usepackage{dsfont}
\usepackage{mathrsfs}

\usepackage{color}
\usepackage{xcolor}
\usepackage{xkeyval}

\usepackage[autostyle=true]{csquotes}

\usepackage[round]{natbib}
\setcitestyle{aysep={}}
\usepackage{url}  %
\usepackage{hyperref}

\usepackage[nolist]{acronym}

\usepackage[capitalise,english,nameinlink]{cleveref}

\usepackage{enumerate}
\usepackage{enumitem}
\setlist[enumerate,1]{label=\arabic*.,}
\newlist{inlineitemize}{enumerate*}{1}
\setlist*[inlineitemize,1]{label=(\roman*),}

\usepackage{graphicx}
\usepackage{epsfig}
\usepackage{subcaption}
\usepackage{float}
\restylefloat{table}
\usepackage{wrapfig}
\usepackage[rightcaption]{sidecap}
\sidecaptionvpos{figure}{c}
\usepackage{epstopdf}
\usepackage{dblfloatfix} %

\usepackage[algo2e,linesnumbered,ruled,commentsnumbered,noline]{algorithm2e}

\usepackage{ifthen} %

\usepackage{comment}
\newboolean{todo}
\setboolean{todo}{True} %
\ifthenelse{\boolean{todo}}{\setlength {\marginparwidth}{2cm}
\usepackage{todonotes} %
}{}

\newcommand{\remarkInternal}[4]{\ifthenelse{\boolean{todo}}{\todo[inline, color=#2, caption={2do}, #3]{\begin{minipage}{\textwidth-4pt}
\emph{Remark #1:}\\#4
\end{minipage}}}{}}

\newcommand{\tdBA}[2][]{\remarkInternal{Bastian}{green!20}{#1}{#2}}

\usepackage{tabularx}
\usepackage{booktabs}       %
\usepackage{makecell}
\usepackage{hhline}
\usepackage{array} %
\usepackage{multirow}

\crefname{paragraph}{paragraph}{paragraphs}
\Crefname{paragraph}{Paragraph}{Paragraphs}
\AtBeginEnvironment{appendices}{
    \crefalias{subsection}{appendix}
    \crefalias{section}{appendix}
    \crefalias{chapter}{appendix}
}
\crefname{appendix}{Appendix}{Appendices}
\crefname{algorithm}{Algorithm}{Algorithms}
\crefname{algocf}{Algorithm}{Algorithms}
\Crefname{algocf}{Algorithm}{Algorithms}



\newcommand{\eg}{e.g.\@\xspace}
\newcommand{\ie}{i.e.\@\xspace}
\newcommand{\wrt}{w.r.t.\@\xspace}

\DeclareMathOperator{\Prob}{\mathrm{P}}
\DeclareMathOperator{\prob}{\mathrm{p}}

\DeclareMathOperator{\E}{\mathsf{E}}

\newcommand{\Eof}{\E\expectarg}
\DeclarePairedDelimiterX{\expectarg}[1]{[}{]}{%
    \ifnum\currentgrouptype=16 \else\begingroup\fi
    \activatebar#1
    \ifnum\currentgrouptype=16 \else\endgroup\fi
}
\newcommand{\innermid}{\nonscript\;\delimsize\vert\nonscript\;}
\newcommand{\activatebar}{%
    \begingroup\lccode`\~=`\|
    \lowercase{\endgroup\let~}\innermid
    \mathcode`|=\string"8000
}

\newcommand{\EofLeft}{\E\expectargleft}
\DeclarePairedDelimiterX{\expectargleft}[1]{[}{.}{%
    \ifnum\currentgrouptype=16 \else\begingroup\fi
    \activatebar#1
    \ifnum\currentgrouptype=16 \else\endgroup\fi
}

\newcommand{\EofRight}{\expectargright}
\DeclarePairedDelimiterX{\expectargright}[1]{.}{]}{%
    \ifnum\currentgrouptype=16 \else\begingroup\fi
    \activatebar#1
    \ifnum\currentgrouptype=16 \else\endgroup\fi
}

\DeclarePairedDelimiterX{\klarg}[1]{(}{)}{%
    \ifnum\currentgrouptype=16 \else\begingroup\fi
    \activatediv#1
    \ifnum\currentgrouptype=16 \else\endgroup\fi
}
\newcommand{\innerdiv}{\nonscript\;\delimsize\Vert\nonscript\;}
\newcommand{\activatediv}{%
    \begingroup\lccode`\~=`\|
    \lowercase{\endgroup\let~}\innerdiv
    \mathcode`|=\string"8000
}

\DeclarePairedDelimiterX{\klargleft}[1]{()}{.}{%
    \ifnum\currentgrouptype=16 \else\begingroup\fi
    \activatediv#1
    \ifnum\currentgrouptype=16 \else\endgroup\fi
}

\DeclarePairedDelimiterX{\klargright}[1]{.}{)}{%
    \ifnum\currentgrouptype=16 \else\begingroup\fi
    \activatediv#1
    \ifnum\currentgrouptype=16 \else\endgroup\fi
}

\DeclareMathOperator{\GamDis}{\mathrm{Gam}}

\DeclareMathOperator{\UniformDis}{\mathrm{Uniform}}
\DeclareMathOperator{\NDis}{\mathcal{N}}

\DeclareMathOperator{\LUDis}{\mathrm{LogUniform}}

\newcommand{\unitvector}{e}
\newcommand{\nSpecies}{d}
\newcommand{\nReactions}{r}
\newcommand{\nSlow}{l}
\newcommand{\nData}{K}

\newcommand\bt{\beta(u,v,t)}
\newcommand\bbt{\beta^{-1}(u,v,t)}
\newcommand\btb{\beta^{-2}(u,v,t)}
\newcommand\bbbt{\beta^{-3}(u,v,t)}
\newcommand\bit{\beta(u+\unitvector_i,v,t)}
\newcommand\bti{\beta(u-\unitvector_i,v,t)}
\newcommand\smt{\tilde{\pi}(u,v,t)}
\newcommand\smti{\tilde{\pi}(u-\unitvector_i,v,t)}
\newcommand\filt{\pi(u,v,t)}
\newcommand\kt{\kappa_j(u,v)}
\newcommand\kiti{\kappa_i(u-\unitvector_i,v)}
\newcommand\kit{\kappa_i(u,v)}
\newcommand\filti{\pi(u-\unitvector_i,v,t)}

\newcommand\bthbb{\beta(u,v,t_{n+1}-h)}

\newcommand\pthn{\prob (y_{n+1:K}\mid u,v,t_{n+1}-h,\phi,y_{1:n})}
\newcommand\pthnn{\prob (y_{n+1:K}, u,v,t_{n+1}-h,\phi,y_{1:n})}
\newcommand\npth{\prob ( u,v,t_{n+1}-h,\phi,y_{1:n})}
\newcommand\pthnnn{\prob (y_{n+1},y_{n+2:K}, u,v,t_{n+1}-h,\phi,y_{1:n})}
\newcommand\pthnnnn{\prob (y_{n+1}\mid y_{n+2:K}, u,v,t_{n+1}-h,\phi,y_{1:n})}
\newcommand\nnpth{\prob (y_{n+2:K}, u,v,t_{n+1}-h,\phi,y_{1:n})}
\newcommand \pthnnnnn{\prob ( y_{n+2:K} \mid u,v,t_{n+1}-h,\phi,y_{1:n})}

\newcommand\bh{\beta(u,v,t-h)}
\newcommand\ph{\prob (y_{n+1:K}\mid u,v,t-h,\phi)}
\newcommand\pph{\prob (y_{n+1:K}\mid u',v',t,u,v,t-h,\phi)}
\newcommand\ppph{\prob (u',v',t\mid u,v,t-h,\phi)}
\newcommand\pppph{\prob (y_{n+1:K}\mid u',v',t,\phi)}
\newcommand\bbh{\beta(u',v',t)}

\jyear{2021}%

\theoremstyle{thmstyleone}%
\newtheorem{theorem}{Theorem}%

\theoremstyle{thmstyletwo}%

\theoremstyle{thmstylethree}%

\begin{document}

    \begin{acronym}
    \acro{mjp}[MJP]{Markov jump process}
    \acro{brn}[BRN]{biochemical reaction network}
    \acro{lna}[LNA]{linear noise  approximation}
    \acroplural{mjp}[MJPs]{Markov jump processes}
    \acroindefinite{mjp}{an}{a}    
    \acro{sde}[SDE]{stochastic differential equation}
    \acro{ode}[ODE]{ordinary differential equation}
    \acroindefinite{ode}{an}{an}
    \acro{ctmc}[CTMC]{continuous-time Markov chain}
    \acro{rtcm}[RTCM]{random time change model}
    \acro{cme}[CME]{chemical master equation}
    \acro{fpe}[FPE]{Fokker-Planck equation}
    \acro{gfpe}[GFPE]{generalized Fokker-Planck equation}
    \acro{hme}[HME]{hybrid master equation}
    \acroindefinite{hme}{an}{a}
    \acro{cle}[CLE]{chemical Langevin equation}
    \acro{smc}[SMC]{sequential Monte Carlo}
    \acroindefinite{smc}{an}{a}
    \acro{sis}[SIS]{sequential importance sampling}
    \acroindefinite{sis}{an}{a}
    \acro{mcmc}[MCMC]{Markov chain Monte Carlo}
    \acroindefinite{mcmc}{an}{a}
    \acro{pmcmc}[PMCMC]{particle Markov chain Monte Carlo}
    \acro{sir}[SIR]{sequential importance resampling}
    \acroindefinite{sir}{an}{a}
    \acro{hmc}[HMC]{Hamiltonian Monte Carlo}
    \acroindefinite{hmc}{an}{a}
    \acro{nuts}[NUTS]{No-U-Turn Sampler}
\end{acronym}

    \title[ ]{Bayesian Inference for Jump-Diffusion Approximations of Biochemical Reaction Networks}

    \author*[1]{\fnm{Derya} \sur{Alt{\i}ntan}}\email{deryaaltintan@hacettepe.edu.tr}
    \author[2]{\fnm{Bastian} \sur{Alt}}\email{bastian.alt@tu-darmstadt.de}
    \author[2,3]{\fnm{Heinz} \sur{Koeppl}}\email{heinz.koeppl@tu-darmstadt.de}

    \affil[1]{\orgdiv{Department of Mathematics}, \orgname{Hacettepe University}, \orgaddress{\city{Ankara} \country{T\"{u}rkiye}}}
    \affil[2]{\orgdiv{Department of Electrical Engineering and Information Technology}, \orgname{Technische Universität Darmstadt}, \orgaddress{\city{Darmstadt} \country{Germany}}}
    \affil[3]{\orgdiv{Department of Biology}, \orgname{Technische Universität Darmstadt}, \orgaddress{\city{Darmstadt} \country{Germany}}}

    \abstract{\Aclp{brn} are an amalgamation of reactions where each reaction represents the interaction of different species.
Generally, these networks exhibit a multi-scale behavior caused by the high variability in reaction rates and abundances of species.
The so-called \emph{jump-diffusion approximation} is a valuable tool in the modeling of such systems.
The approximation is constructed by partitioning the reaction network into a fast and slow subgroup of fast and slow reactions, respectively.
This enables the modeling of the dynamics using a Langevin equation for the fast group, while a \acl{mjp} model is kept for the dynamics of the slow group.
Most often biochemical processes are poorly characterized in terms of parameters and population states.
As a result of this, methods for estimating hidden quantities are of significant interest.
In this paper, we develop a tractable Bayesian inference algorithm based on \acl{mcmc}.
The presented blocked Gibbs particle smoothing algorithm utilizes a \acl{smc} method to estimate the latent states and performs distinct Gibbs steps for the parameters of a biochemical reaction network, by exploiting a jump-diffusion approximation model.

The presented blocked Gibbs sampler is based on the two distinct steps of \emph{state inference} and \emph{parameter inference}.
We estimate states via a continuous-time forward-filtering backward-smoothing procedure in the state inference step.
By utilizing bootstrap particle filtering within a backward-smoothing procedure,  we sample a smoothing trajectory.
For estimating the hidden parameters, we utilize a separate \acl{mcmc} sampler within the Gibbs sampler that uses the path-wise continuous-time representation of the reaction counters.
Finally, the algorithm is numerically evaluated for a partially observed multi-scale birth-death process example.}

    \maketitle

    \section{Introduction}
In general, \acp{brn} contain several species and multiple reaction channels \citep{kam:82,wil:06}, where the copy numbers of the species change in a wide range  and the reactions possess varying time scales.
Traditional approaches, such as pure deterministic models or pure stochastic models, fail to account for this multi-scale nature.
The deterministic approach models the system by using a set of reaction rate equations in the form of \acp{ode} representing the time derivative of the concentrations of species \citep{cornish2013fundamentals}. %
It represents a macroscopic view and therefore fails to model the inherent discrete and stochastic ordinal nature of the underlying \ac{brn}.
As an alternative to the deterministic approach, the stochastic approach models \iac{brn} by using \iac{ctmc}.
This \ac{ctmc} gives a stochastic description for the number of molecules of each species, where the dynamics of the system are fully described by \iac{cme}.
\Iac{cme} is a set of \acp{ode}, possibly of infinite dimension, representing the time derivative of the probability mass function over the number of molecules.
Despite its simplicity, \acp{cme} suffers from the curse of dimensionality, since each state of the system adds an extra differential equation to the corresponding \ac{cme}.
Therefore, the Doob-Gillespie algorithm \citep{doob1945markoff}  and its variants \citep{gill:76,gill:92,gill:07} are used to generate sample paths of the corresponding stochastic process, for a detailed review see, \eg, \citet{karlebach2008modelling}.

Since the computational cost of the Doob-Gillespie algorithm is tremendously demanding for highly reactive systems, hybrid models combining different modeling approaches are needed for the modeling of \acp{brn} exhibiting a multi-scale nature, for a detailed review, see, \eg, \citet{pah:09} and \citet{ah:10}.
A prominent example of a hybrid modeling approach can be found in \citet{hr:02}, where the different modeling approaches are connected in form of a Langevin equation and \iac{ctmc}, see also \citet{dez:16}.
Different simulation strategies to obtain the dynamics of \acp{brn} involving a large number of reactions and species modeled with hybrid methods are proposed in \citet{ssk:06}. 
For an application of these simulation strategies to eukaryotic cell cycles, based on the idea of \citet{hr:02}, see \citet{liu:12}. 
In \citet{kang2019a}, the authors present two hybrid models combining the \ac{ctmc} approach with a stochastic partial differential equation approach.
The work provides a link between the stochastic approach using \acp{ctmc} and the deterministic approach using partial differential equations. 
Different hybrid methods to approximate the solution of the \ac{cme} are proposed in \citet{bak:15} and \citet{mlsh:12}. 
Hybrid simplifications of \acp{brn} using the Kramers-Moyal expansion  \citep{Risken1996} and averaging are analyzed in \citet{cdr:09}, while the convergence analysis of hybrid models based on disparate types of errors is discussed in \citet{ce:18} and \citet{ce:16}. 

\Citet{gak:015} present a jump-diffusion approximation to exploit this multi-scale nature by using the splitting idea used within hybrid models.
The work contributes an error analysis that defines an objective measure to separate the \acp{brn} into different subgroups, which leads to a dynamic separating algorithm.
Based on an error bound the reactions are separated into two groups
\begin{inlineitemize}
    \item a fast group involving species with high copy numbers which is modeled by a diffusion approximation governed by the \ac{cle} \citep{gill:07} and
    \item a slow group involving species with low copy numbers which is modeled by the \ac{ctmc} governed by the \ac{rtcm} \citep{ak:11}.
\end{inlineitemize}
This decomposition results in a path-wise representation of the system under consideration as a combination of a Poissonian \ac{rtcm} and \iac{cle}.
The joint probability density function of the jump-diffusion approximation over the reaction counting process satisfies the \ac{hme}, as proven in \citet{ak:20}, which involves terms from \iac{cme} and \iac{fpe} \citep{paw:67}.
Based on \citet{hwkt:13} and \citet{ak:20} obtain the approximate solution of the  \ac{hme} by constructing moment equations.

A limiting factor in the modeling approaches above is that for real installations of \acp{brn}, it is usually not possible to determine all states and underlying parameters exactly.
Therefore, statistical inference methods that estimate latent states and parameters of \acp{brn} from given observations are needed.
In this regard, Bayesian inference is an essential tool to estimate the latent variables of the system under consideration \citep{gelmanbda04}. 
Unfortunately, the computation of the Bayesian posterior distribution requires in general solving high-dimensional integrals, which for complex models are often computationally intractable, rendering the main drawback for exact Bayesian inference.
A popular approach to overcome this hindrance are \ac{mcmc} methods \citep{bgjm:11}.
They tractably generate samples from the target posterior distribution, which can be used to approximately compute quantities of interest, such as posterior moments, or the posterior density itself using density estimation. 
\Ac{mcmc} methods are offline estimation methods, which generate samples based on the entire observation data set. 
Contrary to that, \ac{smc} methods are an alternative tool, which construct the posterior distribution sequentially, only requiring one observation after another, see, \eg, \citet{cp:20}.

Over the years, these sampling-based strategies have been exploited to estimate unknown states and parameters of \acp{brn}.
For example, in \citet{golightly2006bayesian}, based on the \iac{mcmc} algorithm of \citet{golightly2005bayesian}, the hidden quantities of \acp{brn} are estimated.
Being inspired from \citet{chib2006likelihood}, an efficient \ac{mcmc} method that samples parameters from the posterior distribution conditioned on the Brownian motion of the corresponding \ac{brn} is proposed in \citet{golightly2008bayesian}. In \cite{adh:10}, \ac{pmcmc} methods combining \ac{smc} and \ac{mcmc} techniques are proposed to improve the \ac{mcmc} methods. 
They are utilized to estimate the unknown parameters of \acp{brn} in \citet{gw:11}. 
An inference method for \acp{brn} that uses \ac{pmcmc} together with the \ac{mcmc} technique defined in \citet{geyer1991markov} is presented in \citet{bk:16}.
In \citet{sgg:14}, a \ac{pmcmc} that estimates the hidden quantities of a given \ac{brn} modeled with a hybrid method combining the \ac{lna} and the \ac{ctmc} is proposed. 
A new parallel \ac{mcmc} algorithm based on the \ac{smc} methods to infer the unknown parameters of \acp{brn} is developed in \citet{catanach2020bayesian} while a new Bayesian inference method based on the tensor-train decomposition of the corresponding \ac{cme} of the \acs{brn} under consideration is proposed in \citet{ion2021tensor}. 
We refer the reader to \citet{Schnoerr_2017} and \citet{wil:06} for more details on inference methods for \acp{brn}.

In this work, we propose a Bayesian inference algorithm for jump-diffusion approximations of multi-scale reaction networks.
Based on the works of \citet{gak:015} and \citet{ak:20}, we present a forward model formulation based on the jump-diffusion approximation for \acp{brn} whose probability density function satisfies the \ac{hme}.
To account for partial observability, we present a discrete-time noisy measurement model for the observations of the continuous-time latent chemical reaction network.
To estimate the hidden reaction rates, we consider a full Bayesian setup and quantify the posterior probability of the reaction rates. We give the exact equations for the joint posterior distribution of the latent reaction rates and states given the observations, which are computationally intractable.
Hence, we develop \iac{mcmc} sampler in the form of a blocked Gibbs particle smoother, to infer the latent parameters and states of the system.

The presented Gibbs sampler is divided into two sub-problems of the state inference and the parameter inference.
For the state inference, we sample from the conditional posterior distribution of the states given the parameters and observations by using a forward-filtering backward-smoothing procedure based on a bootstrap filter \citep{Gordon_1993}.
In the parameter inference step, we sample from the full-conditional posterior distribution of the parameters given the observations and the smoothing trajectory generated in the state inference step.
To estimate the fast reaction rate parameters, we use a reparametrization of \citet{chib2006likelihood}, to circumvent mixing problems in the Gibbs sampler and present an equation for the unnormalized density of the full-conditional.
Analogously, to estimate the slow reaction rate parameters, we give an equation for the unnormalized full-conditional density of the slow reaction rate parameters based on Radon-Nikodym derivative of a conditional path measure against a reference measure.
To sample from those unnormalized conditionals, we use an \ac{mcmc} method within the Gibbs sampler.

The rest of the paper is organized as follows:
In \cref{sec:model}, we give a brief summary of the jump-diffusion approximation, the underlying \ac{hme}, together with a characterization for the path measure of the counting processes of the slow reactions. In \cref{sec:posterior_inference}, we present a blocked Gibbs particle smoothing algorithm, namely blocked Gibbs particle smoothing,  and explain the details of the subordinate state and parameter inference steps. In \cref{sec:experiment}, we evaluate the algorithm numerically on an illustrating example and \cref{sec:con} concludes the paper. For an overview of some notational conventions used throughout this paper see \cref{notation}.

    \section{A Partially Observed Jump-Diffusion Model for Reaction Networks}
\label{sec:model}
The traditional stochastic approach describes \acp{brn} as a set of reaction channels.
Each reaction channel in the network describes the interaction between different species.
In this approach, the system's state is represented by the integer-valued copy numbers of species, and the system's dynamics are defined by \iac{ctmc} \citep{ak:11, wil:06}.

We consider a reaction network consisting of $\nReactions \in \mathbb N$ reaction channels $\{R_k\}_{k=1, \dots, \nReactions}$ and $\nSpecies \in \mathbb N$ species $\{S_j\}_{j=1, \dots, \nSpecies}$.
A reaction channel $R_k$, with $k=1, 2, \dots, \nReactions$, can be represented as follows

\begin{equation*}
    R_k : \sum_{j=1}^{\nSpecies} \underline{\mu}_{jk}S_j \stackrel{\phi_{k}}{\longrightarrow} \sum_{j=1}^{\nSpecies} \bar{\mu}_{jk}S_j,
\end{equation*}

where $\phi_k \in \mathbb{R}_{> 0}$ denotes the reaction rate constant of the reaction channel $R_k$.
The non-negative integers $\underline{\mu}_{jk}$ and $\bar{\mu}_{jk}$ are the stoichiometric coefficients.
Here, the coefficients $\underline{\mu}_{jk}$ and $\bar{\mu}_{jk}$ represent the copy number of species $S_j$ used and produced in a single firing of the reaction $R_k$, respectively.
The net change in the copy number of species $S_j$ at the end of a single firing of the reaction $R_k$ is $\mu_{jk}=\bar{\mu}_{jk}-\underline{\mu}_{jk}$ which gives the stoichiometric vector of the reaction  as $\mu_{k}=(\mu_{1k}, \mu_{2k}, \dots, \mu_{dk})^\top$.
We define $X(t)\: \in \mathbb{N}^\nSpecies$ as the state vector of the system at time $t\geq 0$ with the components $X_j(t)$, representing the copy number of the $j$th species $S_j, \: j=1, 2, \dots, \nSpecies$.
If the system is at state $X(t)=x$, then a single firing of reaction channel $R_k$ jumps to the state $x+\mu_k$.
This allows us to describe the state vector $X(t)$ by using reaction counters of the reaction network under consideration.
Let $N(t)=(N_1(t), N_2(t), \dots, N_\nReactions(t))^\top$ denote the vector of reaction counters, where $N_k(t)$ represents the number of firings of the $k$th reaction $R_k$ until time $t>0$, with $k=1, 2, \dots, \nReactions$.
Given $N(t)$, we find a description for the state vector of the multi-scale process $X$ as
\begin{equation}
    X(t)=X(0)+ \sum_{k=1}^{\nReactions} N_k(t) \mu_{k}.
    \label{eq:state_x_y_description}
\end{equation}

In a reaction network, the abundances of species lie in a wide range, from a few copy numbers to millions of copy numbers.
Additionally, the different reaction channels can fire at highly varying speeds. This variability gives rise to a hybrid modeling approach.
The key feature of hybrid models is the separation of reaction channels into different subsets.
In \citet{gak:015}, a jump-diffusion approximation is presented to describe the dynamics of reaction networks, which possess a multi-scale behavior.
The idea of the presented jump-diffusion approximation is to partition the reaction network into two different subgroups,
\begin{inlineitemize}
    \item a subgroup $\mathcal{D}$ of $\nSlow$ slow reactions, \ie, $\vert \mathcal{D} \vert =\nSlow$ and
    \item a subgroup $\mathcal{C}$ of $\nReactions-\nSlow$ fast reactions, \ie, $\vert \mathcal{C}\vert=\nReactions-\nSlow$.
\end{inlineitemize}
Using this partitioning we can simulate the fast reactions by solving \iac{sde}, the \ac{cle}, while samples of the slow reactions can be conducted by the means of \iac{ctmc} simulation.
For the description of the state vector in \cref{eq:state_x_y_description} this partitioning yields
\begin{equation}
    \begin{aligned}
        &X(t)=X(0)+ \sum_{i \in \mathcal{D}} \zeta_i \left( \int_{0}^{t} \gamma_{i} (X (s))\, \mathrm ds\right) \mu_{i} \\
        & + \sum_{j \in \mathcal{C}} \left(\int_{0}^{t} \gamma_{j}(X(s))\, \mathrm ds \right)\, \mu_{j}\\
        &\quad + W_{j} \left( \int_{0}^{t} \gamma_{j} (X(s)) \, \mathrm ds\right) \mu_{j},
    \end{aligned}
    \label{eq:states}
\end{equation}
where $\{ \zeta_i \}_{i \in \mathcal D}$ are independent unit Poisson processes and $\{ W_j \}_{j \in \mathcal C}$ are independent Brownian motions.
Here, $\gamma_k(x)$ represents the propensity function of the reaction $R_k$, $k=1, 2, \dots, r$, satisfying
\begin{equation*}
    \begin{multlined}
        \gamma_k(x)+\frac{o(\Delta t)}{\Delta t}\\
        =\frac{1}{\Delta t}\Prob [X(t+\Delta t)=x+\mu_k \mid X(t)=x],
    \end{multlined}
\end{equation*}
with $ \lim_{\Delta t \rightarrow 0} \frac{o(\Delta t )}{\Delta t}=0$.
Throughout this paper, we assume the law of mass action kinetics for the propensities, \ie,
\begin{equation}
    \begin{aligned}
        &\gamma_k(x)= \phi_k \prod_{j=1}^{\nSpecies}\binom {x_j} {\underline{\mu}_{jk}}, && k=1, 2, \dots, \nReactions.
    \end{aligned}
    \label{eq:propensity_mass_action}
\end{equation}

Given the partitioning of the reactions, we also partition the reaction counters $N(t)$ into sub-components $U(t)=(U_1(t), U_2(t), \dots, U_\nSlow(t))^\top$ and $V(t)=(V_1(t), V_2(t), \dots, V_{\nReactions-\nSlow}(t))^\top$.
Here, $U(t) \in \mathcal{U} \subseteq \mathbb N^{\nSlow}$ is a discrete random variable representing the firing number of the slow reactions until time $t>0$ and $V(t) \in \mathcal{V} \subseteq \mathbb{R}^{\nReactions-\nSlow} $ is a continuous random variable of the number of firings of the fast reactions.
Using this partitioning, we can rewrite \cref{eq:state_x_y_description} as
\begin{equation}
    X(t)=X(0)+ \sum_{i \in \mathcal{D}} U_i(t) \mu_i+ \sum_{j \in \mathcal{C}} V_j(t) \mu_j.
    \label{eq:state_reac_count}
\end{equation}
By comparing \cref{eq:state_reac_count,eq:states} we find a description for the reaction counters as
\begin{align}
    &{}\begin{aligned}
           U_i(t)&= \zeta_i \left( \int_{0}^{t} \gamma_{i} (X (s))\, \mathrm ds\right) \\
           &= \zeta_i \left( \int_{0}^{t} \kappa_{i} (U (s), V(s))\, \mathrm ds\right)
    \end{aligned}\label{eq:dis_count}\\
    &{}\begin{aligned}
           &V_j(t)= \left(\int_{0}^{t} \gamma_{j}(X(s))\, \mathrm ds \right) \\
           &{\phantom{V_j(t)}} + W_{j} \left( \int_{0}^{t} \gamma_{j} (X(s)) \, \mathrm ds\right)\\
           &{\phantom{V_j(t)}}
           \begin{aligned}
               =&\left(\int_{0}^{t} \kappa_{j}(U (s), V(s))\, \mathrm ds \right) \\
               &+ W_{j} \left( \int_{0}^{t} \kappa_{j} (U (s), V(s)) \, \mathrm ds\right),
           \end{aligned}
    \end{aligned}\label{eq:cont_count_u}
\end{align}
with $i \in \mathcal D$, $j \in \mathcal{C}$, and the reaction counter dependent propensities $\kappa_k (u, v)$, $\forall k \in \mathcal{C} \cup \mathcal{D}$, can be computed by plugin \cref{eq:state_reac_count} into the propensities in \cref{eq:propensity_mass_action} as
\begin{equation*}
    \begin{aligned}
        &\kappa_k (u, v) \equiv \gamma_k \left(X(0)+ \sum_{i \in \mathcal{D}} u_i \mu_i+ \sum_{j \in \mathcal{C}} v_j \mu_j\right).
    \end{aligned}
\end{equation*}

The resulting dynamics of the hybrid system modeled by the jump-diffusion approximation can be characterized by the time-point-wise marginal
$
        \prob (u, v, t\mid \mathcal{H}) %
        \coloneqq \partial_{v_1} \partial_{v_2}\dots \partial_{v_{\nReactions-\nSlow}} \Prob (V(t) \leq v, U(t)=u \mid \mathcal{H}),
$
where $\mathcal{H}$ denotes an arbitrary set involving, \eg, reaction rates $\{\phi_k\}_{k=1, 2, \dots, \nReactions}$ and initial values $U(0)=u_0$, $V(0)=v_0$.
The characterization is given by the following theorem.
\begin{theorem}
    \label{thm_hybrid}
    Let $N(t)=(U^\top(t), V^\top(t))^\top$ denote a joint counting process. Here, $U$ represents the discrete random process with realizations
    $u \in \mathcal{U} \subseteq \mathbb{N}^{\nSlow}, $ while $V$ represents the continuous random process with realizations
    $v \:\in \mathcal{V} \subseteq \mathbb{R}^{\nReactions-\nSlow}$.
    The state vector of the multi-scale process $X$ is given by \cref{eq:state_reac_count} and the counting processes $U$ and $V$ satisfy \cref{eq:dis_count,eq:cont_count_u}, respectively.
    Then, the time-point-wise marginal $\prob (u, v, t\mid \mathcal{H})$
    satisfies \iac{gfpe} \citep{paw:67}, specifically, the forward \ac{hme} \citep{ak:20}
    \begin{equation}
        \partial_t \prob (u, v, t \mid \mathcal{H}) = \mathscr{A} \prob (u, v, t \mid \mathcal{H}),
        \label{eq:hme_general}
    \end{equation}
    subject to the given initial condition $U(0)=0$ and $V(0)=0$, \ie, $\prob (u, v, 0 \mid \mathcal{H})= \delta(u)\delta(v)$

    Here, $\mathscr{A}(\cdot)=\mathscr{D} (\cdot)+\mathscr{C} (\cdot)$ is defined by
    \begin{equation*}
        \begin{aligned}
        &\begin{multlined}
             \mathscr{D} \prob (u, v, t \mid \mathcal{H}) = \sum_{i \in \mathcal{D}} \kappa_{i}(u-\unitvector_i, v) \prob (u-\unitvector_i, v, t \mid \mathcal{H})\\
            -\kappa_{i}(u, v)\prob (u, v, t \mid \mathcal{H})
        \end{multlined}\\
        &\begin{multlined}
            \mathscr{C} \prob (u, v, t \mid \mathcal{H})
             = -\sum_{j \in \mathcal{C}} \partial_{v_j} ( \kappa_{j}(u, v) \prob (u, v, t \mid \mathcal{H}))\\
            +\frac{1}{2} \sum_{j \in \mathcal{C}} \partial_{v_j}^{2} ( \kappa_{j}(u, v) \prob (u, v, t \mid \mathcal{H})).
        \end{multlined}
        \end{aligned}
    \end{equation*}

\end{theorem}
A proof for the above theorem can be found in \cref{sec:proof_hme}.

Similarly, there is an analog backward \ac{hme} \citep{kah:21, paw:67} for the density
$
    \prob (\mathcal{H}\mid u, v, t) \coloneqq \prob (\mathcal{H}\mid U(t)=u, V(t)=v),
$
which is given by
\begin{equation*}
    \partial_t \prob (\mathcal{H}\mid u, v, t)= -\mathscr{A}^\dag \prob (\mathcal{H}\mid u, v, t),
\end{equation*}
where the operator $\mathscr{A}^\dag(\cdot)=\mathscr{D}^\dag (\cdot)+\mathscr{C}^\dag(\cdot)$ is characterized by
\begin{equation*}
    \begin{aligned}
    &\begin{aligned}
    \mathscr{D}^\dag \prob (\mathcal{H} \mid u, v, t)
        = \sum_{i \in \mathcal{D}} \kappa_{i}(u, v) (\prob(\mathcal{H}\mid u+\unitvector_i, v, t&) \\
       -\prob (\mathcal{H}\mid u, v, t &) )
    \end{aligned}\\
    &\begin{multlined}
    \mathscr{C}^\dag \prob (\mathcal{H} \mid u, v, t) = \sum_{j \in \mathcal{C}} \kappa_{j}(u, v) \partial_{v_j}  \prob (\mathcal{H} \mid u, v, t)\\
        +\frac{1}{2} \sum_{j \in \mathcal{C}} \kappa_{j}(u, v) \partial_{v_j}^{2}  \prob (\mathcal{H} \mid u, v, t ).
    \end{multlined}
    \end{aligned}
\end{equation*}
Here, $\mathscr{A}$ and $\mathscr{A}^\dag$ are adjoints of each other \wrt the inner product
$\langle \prob , \mathrm{q} \rangle\coloneqq \sum_{u \in \mathcal{U}} \int \prob (u, v, t) \mathrm{q}(u, v, t) \, \mathrm dv$, that is,
\begin{equation*}
    \langle \mathscr{A} \prob , \mathrm{q} \rangle = \langle \prob , \mathscr{A}^\dag \mathrm{q} \rangle.
\end{equation*}

\subsection{A Path-Wise Characterization of the Counting Process}
\label{sec:path_measure}
Often we can also find a path-wise description of a stochastic process, compared to the time-point-wise marginals discussed before.
Here, we give a characterization of the counting process $U$ representing the reaction counters of the slow reactions as in \cref{eq:dis_count} for a given process $V$ representing the reaction counters of the fast reactions  as in \cref{eq:cont_count_u}.

We have $\nSlow$ slow reactions in our system, therefore, the state vector of the process $U$ at time $t \geq 0$ is $U(t)=(U_1(t), U_2(t), \dots, U_{\nSlow}(t) )^\top$, where $U_i(t)$ represents the firing number of the reaction $R_i$, with $i \in \mathcal{D}$, in the time interval $[0, t]$.
Since the Markov chain representation is kept for slow reactions, as mentioned above, the reaction counting process is a Poisson process, \ie,
\begin{equation}
    \begin{aligned}
        &U_i(t) = \zeta_i \left( \int_{0}^{t} \kappa_{i} (U (s), V(s))\, \mathrm ds\right) && i \in \mathcal D,
    \end{aligned}
    \label{eq:state_eq_slow_reaction}
\end{equation}
where $\zeta_i$ represents the independent unit Poisson processes, see, \eg, \citet{ak:11}.
To obtain a description for a density, we compute in \cref{sec:radon_nikodym} the Radon-Nikodym derivative
\begin{equation*}
    \begin{aligned}
        D(u_{[0, T]}) &\coloneqq \frac{\mathrm d \Prob_{U\mid V,\Phi}}{\mathrm d \Prob_{\zeta}}(u_{[0, T]})\\
        &=\frac{\Prob (U_{[0, T]} \in \mathrm d u_{[0, T]}\mid v_{[0, T]}, \phi)}{\Prob (\zeta_{[0, T]} \in \mathrm d u_{[0, T]})}.
    \end{aligned}
\end{equation*}
This Radon-Nikodym derivative between the path measure $\Prob_{U\mid V,\Phi}$ of the stochastic process $U$ given $V$ characterized by \cref{eq:state_eq_slow_reaction} and the path measure $\Prob_{\zeta}$ of the multivariate Poisson process $\zeta(t)=(\zeta_1(t), \zeta_2(t), \dots, \zeta_{\nSlow}(t) )^\top$ yields the following density expression
\begin{equation}
    \begin{split}
        D(u_{[0, T]})=\exp \left(\int_{0}^{T} \sum_{i \in \mathcal{D}} [1-\kappa_i(u(s), v(s))]\, \mathrm d s \right) \\
        \cdot \prod_{i \in \mathcal D} \prod_{j=1}^{u_i(T)} \kappa_{i}
        (u(\tau_{i,j}^{-}), v(\tau_{i,j}^{-})),
    \end{split}
    \label{eq:like_mjp}
\end{equation}
where $\tau_{i,j}^{-}$ is the time right before the $j$th firing time of the $i$th slow reaction $R_i$, $i \in \mathcal D$ and $u_i(T)$ is the corresponding number of firings in the time interval $[0, T]$.
Note that these results can also be extended to the general case comparing two measures of jump-diffusion processes using Girsanov's theorem, see \citet{hanson2007applied} for an accessible introduction, and \citet{cheridito2005equivalent} and \citet{oksendal2005stochastic} for mathematical treatments.

\subsection{Partial Observability}
Finally, in most setups the state $X(t)$ can not be observed directly.
Rather, often only noisy measurements $Y_n \coloneqq Y(t_n)$ of the state $X_n \coloneqq X(t_n)$ at discrete time points $\{t_n\}_{n=1, \dots, \nData}$ are available.
To capture this setup, we model the measurements using a probabilistic model given as
\begin{equation*}
    \begin{aligned}
        &Y_n \mid X_n \sim \prob (y_n \mid x_n), &&n=1, 2, \dots, \nData.
    \end{aligned}
\end{equation*}
    \section{Posterior Inference}
\label{sec:posterior_inference}
Statistical inference aims to estimate unknown quantities of the system from observations.
For this, we consider a time interval $[0, T]$ and resort to a Bayesian approach.
In this setup, the latent quantities are characterized by a conditional probability of
\begin{inlineitemize}
    \item the state path $X_{[0, T]}$ and
    \item the reaction rates $\Phi \coloneqq (\Phi_1, \dots, \Phi_\nReactions)^\top$
\end{inlineitemize}
given the observation data $Y_{1:\nData}$ in the time interval, \ie,
\begin{equation}
    X_{[0, T]}, \Phi \mid Y_{1:\nData} \sim \Prob (\mathrm d x_{[0, T]}, \mathrm d \phi \mid y_{1:\nData}),
    \label{eq:path_measure_general}
\end{equation}
where $\Prob (\mathrm d x_{[0, T]}, \mathrm d \phi \mid y_{1:\nData}) \coloneqq \Prob (X_{[0, T]}\in \mathrm d x_{[0, T]}, \Phi \in \mathrm d \phi \mid Y_{1:\nData}=y_{1:\nData})$.

An equivalent characterization of \cref{eq:path_measure_general} is given by the joint posterior over the firing counters and the reaction rates
\begin{equation}
    \begin{multlined}
        U_{[0, T]}, V_{[0, T]}, \Phi \mid Y_{1:\nData} \\
        \sim \Prob (\mathrm d u_{[0, T]}, \mathrm d v_{[0, T]}, \mathrm d \phi \mid y_{1:\nData}),
        \label{eq:full_posterior}
    \end{multlined}
\end{equation}
as we can easily transform the firing counters $U(t)$ and $V(t)$ into the state $X(t)$ using \cref{eq:state_reac_count}, \ie,
\begin{equation*}
    X(t)=X(0) + \sum_{i \in \mathcal D} U_i(t) \mu_i + \sum_{j \in \mathcal C} V_j(t) \mu_j,
\end{equation*}
where we assume a given initial value $X(0)=x_0$.

For inferring the reaction rates $\Phi$ we place a prior on them, which yields a generative model.
This forward model consists of drawing the reaction rates from the prior distribution
\begin{equation*}
    \Phi \sim \prob (\phi),
\end{equation*}
subsequently simulating the firing counters
\begin{equation*}
    U_{[0, T]}, V_{[0, T]}\mid \Phi \sim \Prob (\mathrm d u_{[0, T]}, \mathrm d v_{[0, T]} \mid \phi)
\end{equation*}
and drawing the observations as
\begin{equation*}
    Y_n \mid U(t_n), V(t_n) \sim \prob (y_n\mid u_n, v_n), \; n=1, \dots, \nData.
\end{equation*}
Here, the measurement density is given by $ \prob (y_n\mid u_n, v_n)=\prob (y_n \mid x_n)$, where the state $x_n$ is computed as in \cref{eq:state_reac_count}, as
\begin{equation*}
    x_n=x_0 + \sum_{i \in \mathcal{D}} u_{n, i} \mu_i+ \sum_{j \in \mathcal{C}} v_{n, j} \mu_j,
\end{equation*}
with the realizations of the counters $U_{n, i}\coloneqq U_{i}(t_n)$ and $V_{n, j}\coloneqq V_{j}(t_n)$, for $i \in \mathcal D$ and $ j \in \mathcal C$.

Given the generative model, the exact posterior distribution in \cref{eq:full_posterior} can be computed as
\begin{equation*}
    \begin{aligned}
        &\Prob (\mathrm d u_{[0, T]}, \mathrm d v_{[0, T]}, \mathrm d \phi \mid y_{1:\nData}) \\
        &= \frac{\prob ( y_{1:\nData} \mid u_{[0, T]}, v_{[0, T]}, \phi) \Prob (\mathrm d u_{[0, T]}, \mathrm d v_{[0, T]}, \mathrm d \phi)}{\prob (y_{1:\nData})}
    \end{aligned}
\end{equation*}
which requires computing the evidence
\begin{equation}
    \begin{aligned}
        &\prob (y_{1:\nData})=\\
        &\int \prob ( y_{1:\nData} \mid u_{[0, T]}, v_{[0, T]}, \phi) \Prob (\mathrm d u_{[0, T]}, \mathrm d v_{[0, T]}, \mathrm d \phi )
    \end{aligned}
    \label{eq:evidence}
\end{equation}
This computation is an intractable problem because it requires computing an integral over the space of all reaction rates $\phi$ and all paths $u_{[0, T]}$ and $v_{[0, T]}$.

Even though the computation of the posterior distribution is intractable it is often useful to characterize the posterior path measure $\Prob (\mathrm d u_{[0, T]}, \mathrm d v_{[0, T]}, \mathrm d \phi \mid y_{1:\nData})$, by its time-point-wise marginal density
\begin{equation*}
    \prob (u, v, t, \phi \mid y_{1:\nData})= \prob (u, v, t \mid \phi, y_{1:\nData}) \prob (\phi \mid y_{1:\nData}).
\end{equation*}
Here, $\prob (\phi \mid y_{1:\nData})$ is the marginal posterior of the parameters and $\prob (u, v, t \mid \phi, y_{1:\nData})$ is
the \emph{smoothing distribution}, see, \eg, \citet{sarkka_2013}, \citet{anderson1983smoothing}, and \citet{kah:21}, which we define as
\begin{equation*}
    \tilde{\pi}(u, v, t) \coloneqq \prob (u, v, t \mid \phi, y_{1:\nData}).
\end{equation*}
The smoothing distribution can be computed utilizing Bayes' rule as
\begin{equation}
    \begin{aligned}
        & \tilde{\pi}(u, v, t)\\
        &=\frac{ \prob  (u, v, t, y_{1:n}, y_{n+1:\nData}\mid \phi)}{ \prob  (y_{1:n}, y_{n+1:\nData} \mid \phi)} \\
        &=\frac{ \prob  (y_{n+1:\nData}\mid u, v, t, \phi, y_{1:n} )}{ \prob  (y_{n+1:\nData} \mid \phi, y_{1:n})} \prob  ( u, v, t\mid \phi, y_{1:n})\\
        &=\frac{ \prob  (y_{n+1:\nData}\mid u, v, t, \phi )}{ \prob  (y_{n+1:\nData} \mid \phi, y_{1:n})} \prob  ( u, v, t \mid \phi, y_{1:n})\\
        &=\tilde Z_{n}^{-1} \beta (u, v, t) \pi (u, v, t).
    \end{aligned}\label{eq:smoothing_bayes_rule}
\end{equation}
The above quantities in \cref{eq:smoothing_bayes_rule} can be identified as, firstly, the \emph{filtering distribution}
\begin{equation*}
    \pi(u, v, t) \coloneqq \prob (u, v, t\mid \phi, y_{1:n}),
\end{equation*}
which is the posterior distribution at time $t$ conditioned on the observations $Y_{1:n}$ received up until that time, \ie, $n = \max \{n'\in \mathbb N \mid t_{n'} \leq t \}$ and the parameters $\phi$.
Secondly, in \cref{eq:smoothing_bayes_rule} the \emph{backward distribution} is
\begin{equation*}
{\beta}(u, v, t)
    \coloneqq \prob  (y_{n+1:\nData}\mid u, v, t, \phi),
\end{equation*}
which is a backward filtering quantity, that is the likelihood of the \enquote{future} observations $Y_{n+1:\nData}$. Finally, a normalizing constant is given by
\begin{equation*}
\begin{aligned}
      \tilde{Z}_n & \coloneqq \prob  (y_{n+1:\nData} \mid \phi, y_{1:n})\\
      &= \sum_{u \in \mathcal{U}} \int \beta(u, v, t) \pi(u, v, t) \, \mathrm dv,
\end{aligned}
\end{equation*}
It can be shown that the filtering distribution $ \pi(u, v, t)$, the backward distribution $ {\beta}(u, v, t)$, as well as the smoothing distribution $\tilde{\pi}(u, v, t)$, can be computed recursively.
Specifically, the time-evolution equation of the filtering distribution between the observation points follows the \ac{hme}, see \cref{eq:hme_general},
\begin{equation*}
    \partial_t \pi(u, v, t) =\mathscr{A} \pi(u, v, t),
\end{equation*}
with initial condition $\pi (u, v, 0 )= \delta(u)\delta(v)$.
The reset conditions at the observation points are given as
\begin{equation*}
    \pi(u, v, t_n)=Z_n^{-1}\prob (y_n \mid u, v)\pi(u, v, t_n^{-}),
\end{equation*}
where we denote by $\pi(u, v, t_n^{-})$ the filtering distribution right before the $n$th observation, \ie,
\begin{equation*}
    \pi(u, v, t_n^{-})=\lim_{t \nearrow t_n} \pi(u, v, t)= \prob (u, v, t_n\mid \phi, y_{1:n-1})
\end{equation*}
and we have the normalization constant
\begin{equation*}
    \begin{aligned}
        Z_n &=\prob (y_n \mid \phi, y_{1:n-1})\\
        &= \sum_{u \in \mathcal{U}} \int \prob (y_n \mid u, v) \pi(u, v, t_n^{-}) \, \mathrm dv,
    \end{aligned}
\end{equation*}
for more details see \cref{sec:filts}.
Similarly, the time derivative \wrt the density of the backward distribution between the observation points satisfies
\begin{equation*}
    \partial_t \bt = -\mathscr{A}^\dag \bt,
\end{equation*}
subject to the the end point condition $\beta(u, v, T)=1$, with the adjoint operator $\mathscr{A}^\dag$.
The backward distribution at the observation points satisfies
\begin{equation*}
    \begin{aligned}
        \beta(u, v, t_{n+1}^{-})=\beta(u, v, t_{n+1}) \prob (y_{n+1}\mid u, v),
    \end{aligned}
\end{equation*}
with
\begin{equation*}
    \beta(u, v, t_{n+1}^{-})=\lim_{t \nearrow t_{n+1}} \beta(u, v, t)
\end{equation*}
for details see \cref{sec:backward}.
Finally, the time derivative \wrt the density of the smoothing distribution is given as follows
\begin{equation*}
    \begin{aligned}
        &\partial_{t} \smt \\
        &=- \sum_{j \in \mathcal{C}} \partial_{v_j} \{ \kt +\partial_{v_j} \log (\bt) \kt \} \\
        &\quad \cdot \smt + \sum_{j \in \mathcal{C}} \partial_{v_j}^2 (\kt \smt ) \\
        &\quad + \sum_{i \in \mathcal{D}} \kiti \smti \frac{ \bt}{\bti} \\
        &\quad - \sum_{i \in \mathcal{D}} \kit \smt \frac{ \bit}{\bt},
    \end{aligned}
\end{equation*}
with initial condition $\tilde{\pi}(u,v,0)=\delta(u)\delta(v)$, for more see \cref{sec:smth}.
Though, the point-wise expressions give us a characterization of the path-wise posterior distribution in form of a density the required calculations are still intractable as in \cref{eq:evidence}.
This is because computing besides the marginal posterior $\prob (\phi \mid y_{1:\nData})$, the time-evolution of the filtering distribution $ \pi(u, v, t)$, the backward distribution $ {\beta}(u, v, t)$, as well as calculating the required normalization constants $\tilde{Z}_n$ all still require to solve high-dimensional integrals and sums over the state variables and rate parameters.

To circumvent computing such intractable integrals, \ac{mcmc} methods \citep{bgjm:11, gelmanbda04, rs:97} are a valuable computational tool for Bayesian statistics.
\Ac{mcmc} methods are widely applied in areas such as engineering \citep{ps:15, wh:12}, epidemics \citep{hmr:13, phi:02}, and biochemistry \citep{vf:19, tn:19}.
They construct a Markov chain, where the stationary distribution is the probability distribution of interest.
Therefore, they can produce samples from the target posterior distribution, without suffering from the curse of dimensionality.
There has been a substantial development of these techniques, including various extensions of the  Metropolis-Hastings algorithm \citep{hastings:70, Metropolis1953}, such as the Metropolis-adjusted Langevin algorithm and \ac{hmc}, see, \eg, \cite{duane1987hybrid, neal2011mcmc}, and extensions like the \acf{nuts} of \cite{hoffman2014no}.
However, these types of acceptance-rejection schemes can be slow if they are naively applied to state space models like the one presented here.
Therefore, often-times a Gibbs sampling scheme, see, \eg, \citet{gelmanbda04} and \citet{gemangeman:84}, is preferable, where first the latent state variables conditioned on all other variables are drawn and subsequently the parameters are sampled conditioned on all other variables.

In this work, we develop a blocked Gibbs particle smoothing scheme to sample from the full posterior distribution in \cref{eq:full_posterior}.
In the presented scheme, we want to alternatingly sample the joint paths $(U_{[0, T]}, V_{[0, T]})$ and the reaction rates $\Phi$ conditioned on each other and the data $Y_{1:\nData}$, \ie,
\begin{equation}
    \begin{aligned}
        & \begin{multlined}
              U_{[0, T]}^{(m)}, V_{[0, T]}^{(m)} \mid \Phi^{(m-1)}, Y_{1:\nData} \\
              \sim \Prob (\mathrm d u_{[0, T]}, \mathrm d v_{[0, T]} \mid y_{1:\nData}, \phi)
        \end{multlined}\\
        &\begin{multlined}
             \Phi^{(m)} \mid U_{[0, T]}^{(m)}, V_{[0, T]}^{(m)}, Y_{1:\nData} \\
             \sim \prob (\phi \mid u_{[0, T]}, v_{[0, T]}, y_{1:\nData}),
        \end{multlined}
    \end{aligned}
\end{equation}
where $m$ denotes the iteration step of the algorithm.
However, note that the path $V_{[0, T]}$ is the solution to \iac{sde}, see, \eg, \citet{ek:86}, as
\begin{equation*}
    \begin{aligned}
        V_j(t)=&\left(\int_{0}^{t} \kappa_{j}(U (s), V(s))\, \mathrm ds \right)\\
        &+ W_{j} \left( \int_{0}^{t} \kappa_{j} (U (s), V(s)) \, \mathrm ds\right)\\
        \Leftrightarrow\, \mathrm d V_j(t)= & \kappa_{j}(U (t), V(t))\, \mathrm dt \\
        &{}+\sqrt{ \kappa_{j}(U(t), V(t))}\, \mathrm d W_j(t).
    \end{aligned}
\end{equation*}
This is problematic as performing Gibbs sampling by alternating sampling between parameters and the solutions of \acp{sde} are known to suffer from convergence issues, see, \eg, \citet{chib2006likelihood} and \citet{golightly2008bayesian}.
This is sometimes termed the Roberts-Stramer critique named after \citet{roberts2001inference}, which first discussed these convergence issues in the context of univariate diffusions.
The problem is that parameters appearing in the dispersion of the \ac{sde} can be deterministically computed using the quadratic variation of the diffusion process.
This leads to a degenerate sampler with a bad mixing behavior, since the conditional density for the parameters is peaked at the value that was previously used to generate the diffusion path.
Therefore, we first split the parameter updates into separate Gibbs steps, \begin{inlineitemize}
\item for slow reaction rate parameters $\{\Phi_i\}_{i \in \mathcal D}$ and
\item the fast reaction rate parameters $\{\Phi_j\}_{j \in \mathcal C}$ involved in the dispersion of the diffusion process.
\end{inlineitemize}
This yields the following blocked Gibbs sampler
\begin{align*}
    &\begin{aligned}
         &U_{[0, T]}^{(m)}, V_{[0, T]}^{(m)} \mid \Phi^{(m-1)}, Y_{1:\nData} \\
         &\quad \sim \Prob (\mathrm d u_{[0, T]}, \mathrm d v_{[0, T]} \mid y_{1:\nData}, \phi)
    \end{aligned}\\
    &\begin{aligned}
         &\{\Phi_i^{(m)}\}_{i \in \mathcal D} \mid U_{[0, T]}^{(m)}, V_{[0, T]}^{(m)}, \{\Phi_j^{(m-1)}\}_{j \in \mathcal C}, Y_{1:\nData} \\
         &\quad \sim \prob ( \{\phi_i\}_{i \in \mathcal{D}} \mid u_{[0, T]}, v_{[0, T]}, \{\phi_j\}_{j \in \mathcal{C}}, y_{1:\nData})\\
         &\{\Phi_j^{(m)}\}_{j \in \mathcal C} \mid U_{[0, T]}^{(m)}, V_{[0, T]}^{(m)}, \{\Phi_i^{(m)}\}_{i \in \mathcal D}, Y_{1:\nData} \\
         &\quad \sim \prob ( \{\phi_j\}_{j \in \mathcal{C}} \mid u_{[0, T]}, v_{[0, T]}, \{\phi_i\}_{i \in \mathcal{D}}, y_{1:\nData}).
    \end{aligned}
\end{align*}
Next, we use a reparameterization of \citet{chib2006likelihood} for the sampler.
The idea is to sample the conditional Brownian motion $W_{[0, T]}$ instead of the conditional diffusion path $V_{[0, T]}$, which is known to alleviate the convergence issues.
For this we use the one-to-one correspondence between the Brownian motions $\{W_j(t)\}$ and the counters $\{V_j(t)\}$ as
\begin{equation}
    \begin{aligned}
        &\mathrm d V_j(t)\\
        &{}=\kappa_{j}(U (t), V(t))\, \mathrm dt +\sqrt{ \kappa_{j}(U(t), V(t))}\mathrm d W_j(t)
    \end{aligned}\label{eq:sde_forward_param}
\end{equation}
and consequently, we have
\begin{equation}
    \begin{aligned}
        \mathrm d W_j(t)=\frac{\mathrm d V_j(t)-\kappa_{j}(U (t), V(t))\, \mathrm dt}{\sqrt{ \kappa_{j}(U(t), V(t))}}.
    \end{aligned}\label{eq:sde_inverse_param}
\end{equation}
Therefore, we build a non-degenerate version of the Gibbs sampler by performing the following update scheme

\begin{align}
    &\begin{aligned}
         &U_{[0, T]}^{(m)}, V_{[0, T]}^{(m)} \mid \Phi^{(m-1)}, Y_{1:\nData} \\
         &\quad \sim \Prob (\mathrm d u_{[0, T]}, \mathrm d v_{[0, T]} \mid y_{1:\nData}, \phi)
    \end{aligned}\label{eq:gibbs_step_state} \\
    &\begin{aligned}
         &\{\Phi_i^{(m)}\}_{i \in \mathcal D} \mid U_{[0, T]}^{(m)}, V_{[0, T]}^{(m)}, \{\Phi_j^{(m-1)}\}_{j \in \mathcal C}, Y_{1:\nData} \\
         &\quad \sim \prob ( \{\phi_i\}_{i \in \mathcal{D}} \mid u_{[0, T]}, v_{[0, T]}, \{\phi_j\}_{j \in \mathcal{C}}, y_{1:\nData})
    \end{aligned}\label{eq:gibbs_step_param_slow}\\
    &\mathrm d W_j^{(m)}(t)=\frac{\mathrm d V_j^{(m)}(t)-\kappa_{j}(U^{(m)} (t), V^{(m)}(t))\, \mathrm dt}{\sqrt{ \kappa_{j}(U^{(m)}(t), V^{(m)}(t))}}\label{eq:repara_chib}\\
    &\begin{aligned}
         &\{\Phi_j^{(m)}\}_{j \in \mathcal C} \mid U_{[0, T]}^{(m)}, W_{[0, T]}^{(m)}, \{\Phi_i^{(m)}\}_{i \in \mathcal D}, Y_{1:\nData} \\
         &\quad \sim \prob ( \{\phi_j\}_{j \in \mathcal{C}} \mid u_{[0, T]}, w_{[0, T]}, \{\phi_i\}_{i \in \mathcal{D}}, y_{1:\nData})
    \end{aligned}\label{eq:gibbs_step_param_fast}
\end{align}
Here, the first step in \cref{eq:gibbs_step_state} yields a sample from the conditional posterior of the reaction counters given the parameters and the observation data, which corresponds to the problem of \emph{state inference}.
In this blocked Gibbs step, we draw a \emph{smoothing trajectory} by using a forward-filtering backward-smoothing procedure whose details are discussed \cref{sec:state_infer_section}.
In the Gibbs step for the parameters in \cref{eq:gibbs_step_param_slow,eq:gibbs_step_param_fast}, discussed in \cref{sec:param_inference}, a sample from the conditional distribution of the parameters is drawn, which we refer to as \emph{parameter inference}.
For the parameter inference of the fast-reaction rate parameters, we reparameterize the distribution in terms of the posterior Brownian motion in \cref{eq:repara_chib} to alleviate the mixing problems in the naive Gibbs sampler.
Note that, we compute the propensities $\kappa_{j}(U^{(m)} (t), V^{(m)}(t))$ in \cref{eq:repara_chib} \wrt the parameters $\Phi^{(m-1)}$.

\subsection{State Inference}
\label{sec:state_infer_section}
As noted before, the main drawback of Bayesian inference, in general, is the presence of intractable sums and integrals.
There are two widely used methods in state space models to circumvent these intractabilities, which are Kalman filter-based methods and \ac{smc} methods.

Kalman filtering \citep{kalman1961new} is utilized to estimate hidden states of linear systems with Gaussian noise.
Over the years different variants of it to infer the hidden states of  more complicated systems have been proposed, for a detailed review, see \citet{khodarahmi2022review},\citet{afshari2017gaussian}.
Unlike Kalman filter-based methods, \ac{smc} methods can be applied to nonlinear state space models with non-Gaussian noise.
\Ac{smc} methods are a combination of \ac{sis} methods and resampling methods, see, \eg, \citet{cgm:07}, \citet{ dj:11}, and \citet{sarkka_2013}.
They are based on the idea of sequentially approximating the posterior distribution by a set of particles.
These particles are distributed using importance weights and a resampling method.
Hence, another common name for \ac{smc} methods is particle filtering, see \eg, \citet{cp:20}, \citet{dfg:01}, \citet{spe:16}.

In this work, for generating a full trajectory from the conditional distribution $\Prob ( \mathrm d u_{[0, T]}, \mathrm d v_{[0, T]} \mid y_{1:\nData}, \phi)$, we utilize a forward-filtering backward-smoothing procedure, see, \eg, \citet{dj:11},  \citet{gdw:04},  \citet{hk:98}, and \citet{or:11}.
The first idea of this procedure is to approximate the target filtering distribution $\pi(u,v,t)$.
In this \emph{forward-filtering step}, the filtering distribution is approximated by an empirical distribution, which is obtained utilizing \iac{smc} method.
Second, in the \emph{backward-smoothing step} we sample from an empirical approximation of the conditional path measure $\Prob ( \mathrm d u_{[0, T]}, \mathrm d v_{[0, T]} \mid y_{1:\nData}, \phi)$.
This empirical distribution is generated backwardly by re-sampling the particles generated by the \ac{smc} method.
Next, we describe these steps in detail.

\subsubsection{Forward-filtering and the Bootstrap Filter}
\label{sec:bootstrap}
In the forward-filtering step of our method, we use \iac{smc} method, a \emph{bootstrap filter}, to build the filtering distribution.
We aim to approximate the filtering distribution $\pi(u,v,t)$, by using an importance sampling method.
For this, we generate samples or \emph{particles} from a \emph{proposal distribution}.
The relation between the target posterior distribution and the proposal distribution are given by the importance weights which are used to obtain an empirical estimate for the target filtering distribution.

We use a \emph{bootstrap filter} \citep{Gordon_1993, sarkka_2013} that uses the prior distribution between the observations as the proposal distribution, \ie,
\begin{equation*}
\begin{multlined}
        U_{[t_{n-1}, t_n]}^{(i)}, V_{[t_{n-1}, t_n]}^{(i)} \mid U(t_{n-1}), V(t_{n-1}), \Phi \\
        \sim \Prob (\mathrm d u_{[t_{n-1}, t_n]},\mathrm d v_{[t_{n-1}, t_n]}\mid u(t_{n-1}), v(t_{n-1}), \phi)
\end{multlined}
\end{equation*}
where $i=1, \dots, M$ is the particle index.
Sampling from this distribution is easy, as we can generate a sample by simulating the system in \cref{eq:state_reac_count,eq:dis_count,eq:cont_count_u}.
The importance weight for the $i$th particle at time point $t_n$ can be computed recursively as
\begin{equation}
    \Gamma_{n}^{(i)} \propto \prob (y_n \mid u_n^{(i)}, v_n^{(i)}) \Gamma_{n-1}^{(i)}.
    \label{eq:particle_weight_recursion}
\end{equation}
This yields an empirical approximation for the filtering distribution as
\begin{equation*}
    \pi(u,v,t) \approx \sum_{i=1}^M \Gamma_n^{(i)} \delta(U^{(i)}(t)-u)\delta(V^{(i)}(t)-v),
\end{equation*}
where $n = \max \{n'\in \mathbb N \mid t_{n'} \leq t \}$.
Additionally, to circumvent particle degeneracy, we perform a resampling procedure, systematic resampling, at the observation time points.
The details of the bootstrap filter are explained in \cref{sec:bootstrap_appendix}, where we explain the \emph{initialization}, \emph{importance resampling}, and \emph{selection} steps.
For more details on particle filters and smoothers, in general, we refer the reader to \citet{ddj:06}, \citet{dj:11}, \citet{spe:16}, and \citet{sarkka_2013}.

\subsubsection{Backward Smoothing}
In the backward-smoothing step, we use \iac{sir} particle smoothing strategy \citep{dj:11, kit:96, sarkka_2013}.
We refer to \cref{sec:backward smoothing appendix} for details and derivations.
For \ac{sir} particle smoothing, we store filtered particles from the forward-filtering step and use them to obtain an empirical approximation of the conditional path measure $\Prob ( \mathrm d u_{[0, T]}, \mathrm d v_{[0, T]} \mid y_{1:\nData}, \phi)$.
Subsequently, our goal is to generate a sample from this conditional distribution.

To achieve this goal, we store the particle trajectories $\{U_{[0, T]}^{(i)}, V_{[0, T]}^{(i)}\}_{i=1}^M$ obtained from the bootstrap filter.
These particles can be interpreted as importance samples of the conditional path measure $\Prob ( \mathrm d u_{[0, T]}, \mathrm d v_{[0, T]} \mid y_{1:\nData}, \phi)$.
It turns out that for \ac{sir} particle smoothing the smoothing weights $\{\tilde{\Gamma}^{(i)}\}_{i=1}^M$ correspond to the last weights of the filtering distribution, i.e.,
\begin{equation*}
        \tilde{\Gamma}^{(i)} = {\Gamma}^{(i)}_K.
\end{equation*}
Hence, an approximation for the sought-after conditional path measure can be obtained via the following particle approximation
\begin{equation*}
    \begin{aligned}
        &\Prob ( \mathrm d u_{[0, T]}, \mathrm d v_{[0, T]} \mid \phi, y_{1:K}) \\
        &\approx \sum_{i=1}^M {\Gamma}^{(i)}_K \delta_{U_{[0, T]}^{(i)}}(\mathrm d u_{[0, T]})\delta_{V_{[0, T]}^{(i)}}(\mathrm d v_{[0, T]}).
    \end{aligned}
\end{equation*}
A sample from this empirical distribution is easily generated, as
\begin{equation}
    \begin{aligned}
        &U_{[0, T]}, V_{[0, T]} \mid \Phi, Y_{1:K} \\
        &\sim \sum_{i=1}^M {\Gamma}^{(i)}_K \delta_{U_{[0, T]}^{(i)}}(\mathrm d u_{[0, T]})\delta_{V_{[0, T]}^{(i)}}(\mathrm d v_{[0, T]}),
    \end{aligned}
    \label{eq:smoothing_emp_sampled}
\end{equation}
implies that the $i$th particle $(U_{[0, T]}^{(i)}, V_{[0, T]}^{(i)})$ is sampled with probability ${\Gamma}^{(i)}_K$.

\begin{figure*}%
    \centering
    \includegraphics{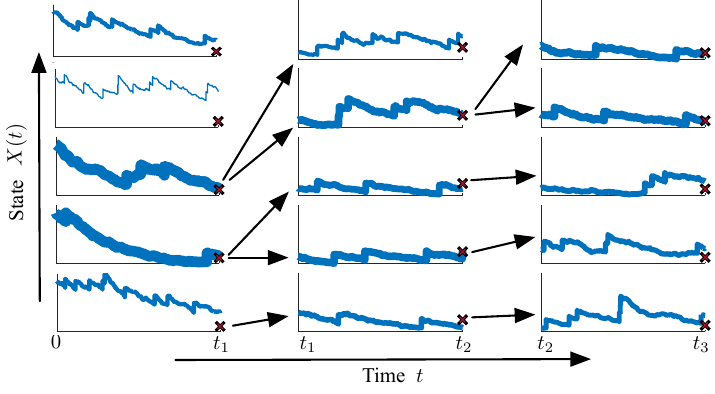}
    \caption{Illustration of the bootstrap filter for a jump-diffusion approximation of a reaction network.
    The figure shows $M=5$ state particles $\{X_{[t_{n-1},t_n]}^{(i)}\}$ for $K=3$ observations, where $i=1,\dots,M$ and $n=1,\dots,K$.
    The counters $U(t)$ and $V(t)$ are converted into the state variable via \cref{eq:state_reac_count}.
    The rows of the figure correspond to the particle index $i$, while the columns are the observation indices $n$.
    The observations are given as red crosses at the time points $t_1<t_2<t_3$.
    The line width denotes the particle weight as in \cref{eq:particle_weight_recursion}.
    Arrows denote the particle selection phase after resampling. Therefore, particles with a high weight (high line width) are replicated, while others are eliminated, for more details see \cref{sec:bootstrap_appendix}.}\label{fig:bfig}
\end{figure*}

\begin{figure*}%
    \begin{subfigure}[b]{0.329\textwidth}
    \centering
        \includegraphics[width=\textwidth]{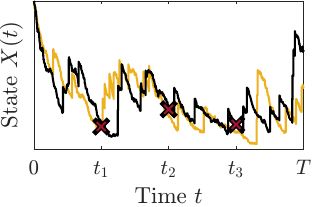}
        \caption{}\label{fig:state_space1}
    \end{subfigure}
    \hfill
    \begin{subfigure}[b]{0.329\textwidth}
    \centering
        \includegraphics[width=\textwidth]{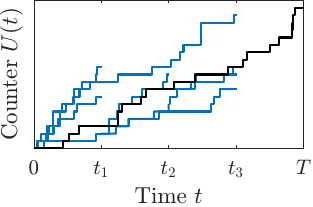}
        \caption{}\label{fig:joint_cme1}
    \end{subfigure}
    \hfill
    \begin{subfigure}[b]{0.329\textwidth}
    \centering
        \includegraphics[width=\textwidth]{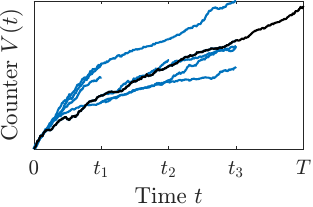}
        \caption{}\label{fig:joint_hme1}
    \end{subfigure}
    \caption{Illustration of the backward smoothing procedure.
    The ground truth latent state trajectory $X(t)$ (yellow line) together with the observations (red crosses) and a smoothing trajectory (black line) are depicted in (\subref{fig:state_space1}).
    A smoothing trajectory (black line) of a discrete reaction counter $U(t)$ and a continuous reaction counter $V(t)$ are shown in (\subref{fig:joint_cme1}) and (\subref{fig:joint_hme1}), respectively.
    The particles of the bootstrap filter (blue line) represent an empirical distribution for the conditional path measure, see \cref{eq:smoothing_emp_sampled}.}\label{fig:bootstatefig}
\end{figure*}

Illustrations of the forward-filtering step and the backward-smoothing are depicted in \cref{fig:bootstatefig,fig:bfig}, respectively.
In \cref{fig:bfig} we show  for a small number of particles for the state $X(t)$ and the observations $Y_{1:\nData}$ an illustration of the bootstrap filter.
The corresponding true latent state trajectory is depicted in \cref{fig:state_space1}.
\Cref{fig:bootstatefig} provides an intuition for the smoothing procedure, where we show in \cref{fig:joint_cme1,fig:joint_hme1} the filter particles of two reaction counters $U(t) \in \mathbb N$ and $V(t) \in \mathbb R$ together with one backward smoothing trajectory.
The backward trajectory is selected according to the empirical distribution in \cref{eq:smoothing_emp_sampled}.
The corresponding smoothing trajectory sample for the state is depicted in \cref{fig:state_space1}.

\subsection{Parameter Inference}\label{sec:param_inference}
Having presented a solution to sampling from the full conditional of the state variables as in \cref{eq:gibbs_step_state}, we present next a method to sample from the full conditionals as in \cref{eq:gibbs_step_param_slow,eq:gibbs_step_param_fast}.
Therefore, we sample from the conditionals $\prob ( \{\phi_i\}_{i \in \mathcal{D}} \mid u_{[0, T]}, v_{[0, T]}, \{\phi_j\}_{j \in \mathcal{C}}, y_{1:\nData})$ and $\prob ( \{\phi_j\}_{j \in \mathcal{C}} \mid u_{[0, T]}, w_{[0, T]}, \{\phi_i\}_{i \in \mathcal{D}}, y_{1:\nData})$ of the slow and fast rate parameters, respectively.

Since computing these conditionals requires in general computing intractable integrals over the parameter space, we next present expressions for the respective unnormalized conditionals.
These unnormalized density expressions can be used by \iac{mcmc} method like, \eg, the Metropolis-Hastings algorithm or \ac{hmc}, to yield a Metropolis-within-Gibbs sampling type scheme.

\subsubsection{Estimating the Slow Reaction Rate Parameters}
\label{sec:ap_slow}
First, we want to estimate the reaction rates of the slow reactions, \ie, $\{\Phi_i\}_{i \in \mathcal{D}}$.
We can find an unnormalized expression for the full conditional $\prob ( \{\phi_i\}_{i \in \mathcal{D}} \mid u_{[0, T]}, v_{[0, T]}, \{\phi_j\}_{j \in \mathcal{C}}, y_{1:\nData})$ of the slow reactions.
For this, we exploit an expression proportional to the path-likelihood
$
   \Prob_{U\mid V,\Phi}(\mathrm d u_{[0, T]})  \coloneqq \Prob (\mathrm d u_{[0, T]} \mid v_{[0, T]}, \{\phi_i\}_{i \in \mathcal{D}}),
$
for which we use the path measure of the discrete counting process whose details are given in \cref{sec:path_measure}.
This yields the following relation
\begin{equation*}
    \begin{aligned}
        &\prob ( \{\phi_i\}_{i \in \mathcal{D}} \mid u_{[0, T]}, v_{[0, T]}, \{\phi_j\}_{j \in \mathcal{C}}, y_{1:\nData})\\
        &\quad \propto \frac{\Prob (\mathrm d u_{[0, T]} \mid v_{[0, T]}, \{\phi_i\}_{i \in \mathcal{D}})}{\Prob_{\zeta}(\mathrm d u_{[0, T]})} \prob (\phi)\\
        &\qquad= \frac{\Prob_{U\mid V,\Phi}(\mathrm d u_{[0, T]})}{\Prob_{\zeta}(\mathrm d u_{[0, T]})} \prob (\phi)= D(u_{[0, T]}) \prob (\phi),
    \end{aligned}
\end{equation*}
where $\Prob_{\zeta}(\mathrm d u_{[0, T]}) \coloneqq \Prob (\zeta_{[0, T]}\in \mathrm d u_{[0, T]})$ is the path measure of the multivariate standard-Poisson process $\zeta$ and
$D(u_{[0, T]}) \coloneqq \frac{\mathrm d \Prob_{U\mid V,\Phi}}{\mathrm d \Prob_{\zeta}}(u_{[0, T]}) \equiv \frac{\Prob_{U\mid V,\Phi}(d u_{[0, T]})}{\Prob_{\zeta}(\mathrm d u_{[0, T]})}$ denotes the Radon-Nikodym derivative between the path-likelihood $\Prob_{U\mid V,\Phi}(\mathrm d u_{[0, T]})$ and the path measure $\Prob_{\zeta}(\mathrm d u_{[0, T]})$, see \cref{eq:like_mjp}.
Hence, using the expression for $D(u_{[0, T]})$ in \cref{eq:like_mjp}, we can generate a sample $\{\Phi_j\}_{j \in \mathcal C}$ of the full conditional in \cref{eq:gibbs_step_param_fast} using the unnormalized density as
\begin{equation}
    \begin{aligned}
        &\prob ( \{\phi_i\}_{i \in \mathcal{D}} \mid u_{[0, T]}, v_{[0, T]}, \{\phi_j\}_{j \in \mathcal{C}}, y_{1:\nData})\\
        &\qquad \propto \exp \left(-\int_{0}^{T} \sum_{i \in \mathcal{D}} \kappa_i(u(s), v(s))\, \mathrm d s \right) \\
        &\qquad \quad \cdot \left(\prod_{i \in \mathcal D} \prod_{j=1}^{u_i(T)} \kappa_{i}
        (u(\tau_{i,j}^{-}), v(\tau_{i,j}^{-}))\right) p(\phi).
    \end{aligned}
    \label{eq:full_conditional_slow}
\end{equation}
Even though computing the normalization constant in \cref{eq:full_conditional_slow} involves in general an intractable integral over the parameters, we can computationally efficiently sample from the expression using the  Metropolis-Hastings algorithm, \ac{hmc} or extensions like \ac{nuts}.

\subsubsection{Estimating the Fast Reaction Rate Parameters}
\label{sec:ap_fast}
Second, we estimate the reaction rates of the fast reactions, \ie, $\{\Phi_j\}_{j \in \mathcal{C}}$.
Using the model structure, we have the following expression for the unnormalized conditional distribution
\begin{equation*}
    \begin{aligned}
        &\prob ( \{\phi_j\}_{j \in \mathcal{C}} \mid u_{[0, T]}, w_{[0, T]}, \{\phi_i\}_{i \in \mathcal{D}}, y_{1:\nData}) \\
        &\propto \prob (y_{1:\nData} \mid u_{[0, T]}, w_{[0, T]}, \phi) \prob ( \phi).
    \end{aligned}
\end{equation*}
The likelihood can be computed as
\begin{equation*}
    \begin{aligned}
        \prob (y_{1:\nData} \mid u_{[0, T]}, w_{[0, T]}, \phi) = \prod_{n=1}^\nData \prob (y_n \mid x_n),
    \end{aligned}
\end{equation*}
where we compute the state $x_n$ using
\begin{equation*}
    x_n =x_0 + \sum_{i \in \mathcal{D}} u_i(t_n) \mu_i+ \sum_{j \in \mathcal{C}} v_j(t_n) \mu_j,
\end{equation*}
with
\begin{equation}
    \begin{aligned}
        &\mathrm d v_j(t)= \\
        &\quad \kappa_{j}(u(t), v(t))\, \mathrm d t +\sqrt{ \kappa_{j}(u(t), v(t))}\, \mathrm d w_j(t). %
    \end{aligned}
    \label{eq:v_reparam_conditional_fast}
\end{equation}
Hence, we can sample from the full conditional of $\{\Phi_j\}_{j \in \mathcal C}$ in \cref{eq:gibbs_step_param_fast} using the unnormalized density
\begin{equation}
    \begin{aligned}
        &\prob ( \{\phi_j\}_{j \in \mathcal{C}} \mid u_{[0, T]}, w_{[0, T]}, \{\phi_i\}_{i \in \mathcal{D}}, y_{1:\nData}) \\
        &\propto \left(\prod_{n=1}^\nData \prob(y_n \mid x_n) \right) \prob(\phi).
    \end{aligned}
    \label{eq:full_conditional_fast}
\end{equation}

This concludes the presentation of the proposed blocked Gibbs particle smoother. A pseudo-code summarizing the sampler is given by \cref{alg:mcmc_sampler}.
\begin{algorithm2e*}
    \DontPrintSemicolon
    \SetKwInOut{Input}{input}\SetKwInOut{Output}{output}
    \Input{$Y_{1:\nData}$: Observation data; $\Phi^{(0)}$: Initial rate parameters; $L$: number of Gibbs samples}
    \Output{Posterior samples $\{U_{[0, T]}^{(m)}, V_{[0, T]}^{(m)}, \Phi^{(m)}\}_{m=1}^L$}

    \For{$m=1$ \KwTo $L$}{
        Sample a smoothing path as in \cref{eq:smoothing_emp_sampled}, by \ac{sir} particle smoothing, see \cref{sec:state_infer_section}, \ie, sample
        \begin{equation*}
        \begin{aligned}
         U_{[0, T]}^{(m)}, V_{[0, T]}^{(m)} \mid \Phi^{(m-1)}, Y_{1:\nData}
         \sim \Prob (\mathrm d u_{[0, T]}, \mathrm d v_{[0, T]} \mid y_{1:\nData}, \phi).
        \end{aligned}
        \end{equation*}

        Sample the slow reaction rate parameters by drawing from \cref{eq:full_conditional_slow}, see \cref{sec:ap_slow}, \ie, sample
        \begin{equation*}
              \begin{aligned}
         \{\Phi_i^{(m)}\}_{i \in \mathcal D} \mid U_{[0, T]}^{(m)}, V_{[0, T]}^{(m)}, \{\Phi_j^{(m-1)}\}_{j \in \mathcal C}, Y_{1:\nData}
         \sim \prob ( \{\phi_i\}_{i \in \mathcal{D}} \mid u_{[0, T]}, v_{[0, T]}, \{\phi_j\}_{j \in \mathcal{C}}, y_{1:\nData}).
        \end{aligned}
        \end{equation*}\;

        Compute the conditional Brownian motions as
        \begin{equation*}
        \begin{aligned}
        &\mathrm d W_j^{(m)}(t)=\frac{\mathrm d V_j^{(m)}(t)-\kappa_{j}(U^{(m)} (t), V^{(m)}(t))\, \mathrm dt}{\sqrt{ \kappa_{j}(U^{(m)}(t), V^{(m)}(t))}}, &&\forall j \in \mathcal C.
        \end{aligned}
        \end{equation*}\;

        Sample the fast reaction rate parameters by drawing from \cref{eq:full_conditional_fast}, see \cref{sec:ap_fast}, \ie, sample
        \begin{equation*}
            \begin{aligned}
             \{\Phi_j^{(m)}\}_{j \in \mathcal C} \mid U_{[0, T]}^{(m)}, W_{[0, T]}^{(m)}, \{\Phi_i^{(m)}\}_{i \in \mathcal D}, Y_{1:\nData}
             \sim \prob ( \{\phi_j\}_{j \in \mathcal{C}} \mid u_{[0, T]}, w_{[0, T]}, \{\phi_i\}_{i \in \mathcal{D}}, y_{1:\nData}).
            \end{aligned}
        \end{equation*}\;
   }
    \caption{Blocked Gibbs Particle Smoothing}\label{alg:mcmc_sampler}
\end{algorithm2e*}
    \section{A Multi-Scale Birth-Death Process Experiment}
\label{sec:experiment}

\begin{figure}
    \centering
    \includegraphics{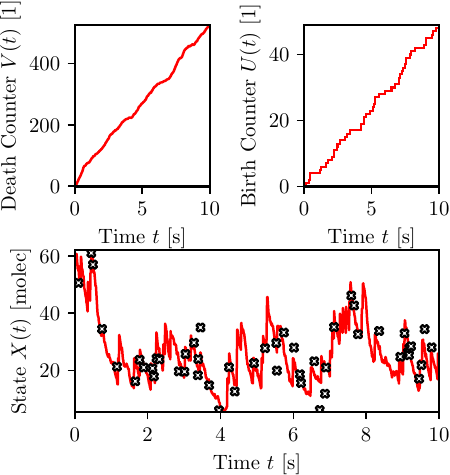}
    \caption{Illustration of the ground-truth realization.
    The trajectories correspond to a static partitioning into slow and fast reaction channels.
    Upper Left Panel: Realization of the reaction counter $U$, modeled as \iac{ctmc}, which corresponds to the dynamics of the reaction counter of the slow reaction $R_2$ given in \cref{eq:count_u}.
    Upper Right Panel: Realization of the reaction counter $V$, modeled using a diffusion approximation, which corresponds to the dynamics of the reaction counter of the fast reaction $R_1$ given in \cref{eq:count_v}.
    Lower Panel: The corresponding realization of the state $X$ as given in \cref{eq:state_ap} and the discrete-time observations $Y_{1:\nData}$ as in \cref{eq:observation_model_exp} depicted as crosses.}
    \label{fig:gt_state_traj}
\end{figure}

In the following, we apply our algorithm to an illustrative example.
We consider a birth-death reaction system with two reactions of the form.
\begin{equation}
    R_1: \rho S \stackrel{\phi_{1}}{\longrightarrow} \emptyset, \quad R_2: \emptyset \stackrel{\phi_{2}}{\longrightarrow} \eta S, \label{eq:bd_example}
\end{equation}
with stoichiometries $\rho \in \mathbb{N}$ and $\eta \in \mathbb{N}$.
In this example, $R_1$ is considered to be a fast reaction and is therefore modeled by a diffusion approximation,
while a discrete state Markov chain updating scheme is kept for the slow reaction $R_2$.
Hence, the sets of fast and slow reactions, the stoichiometries, and the respective change vectors are given by
\begin{equation*}
    \begin{gathered}
        \begin{aligned}
            &\mathcal{C}=\{1\}, &&\mathcal{D}&=\{2\},
        \end{aligned}\\
        \begin{aligned}
            &\underline{\mu}_1=\rho, &&\bar{\mu}_1=0\, \unit{\molec}, &&&\mu_1=-\rho, \\
            &\underline{\mu}_2=0\, \unit{\molec}, &&\bar{\mu}_2= \eta, &&&\mu_2=\eta,
        \end{aligned}
    \end{gathered}
\end{equation*}
where we assume a substrate stoichiometry of $\rho = 1\, \unit{\molec}$ and product stoicheometry of $\eta=10\, \unit{\molec}$.
The system's state vector at time $t \geq 0$ is represented by $X(t) \in \mathbb{R}$.
We divide the reaction counters of the system into two groups, \ie, $N(t)=(U(t), V(t))^\top$, with $U(t) \in \mathbb Z_{\geq 0}$ and $V(t) \in \mathbb R$ representing the firing number of slow and fast reactions until time $t >0$, respectively.
This yields the state vector of the system as
\begin{equation}
    \label{eq:state_ap}
    X(t)=X(0)+\eta U(t) - \rho V(t),
\end{equation}
Where we assume that the state of the system is deterministically initialized as $X(0)=60\, \unit{\molec}$.
The corresponding reaction counters of the system obey the following equations.
\begin{align}
    &\begin{aligned}
         V(t)= &\int_{0}^{t} \kappa_{1}(U(s), V(s))\, \mathrm d s \\
         &{}+ W \left( \int_{0}^{t} \kappa_{1}(U(s), V(s))\, \mathrm d s \right),
    \end{aligned}\label{eq:count_v} \\
    &\begin{aligned}
         U(t) = \zeta \left( \int_{0}^{t} \kappa_{2}(U(s), V(s)) \, \mathrm d s \right),
    \end{aligned} \label{eq:count_u}
\end{align}
and by definition we have $U(0)=V(0)=0$.
The propensity functions above to follow the law of mass action kinetics as
\begin{equation}
    \label{eq:prop}
    \begin{aligned}
        &\begin{multlined}
             \kappa_1(u, v) = \gamma_1(X(0)+\eta u- \rho v)\\=\phi_1 \frac{X(0)+\eta u- \rho v}{\rho},
        \end{multlined}\\
        &\begin{multlined}
             \kappa_2(u, v) = \gamma_2(X(0)+\eta u- \rho v)=\phi_2,
        \end{multlined}
    \end{aligned}
\end{equation}
and the latent rates are set to $\phi_1=2\, \unit{\per\second}$ and $\phi_2=4\, \unit{\per\second}$, hence $\phi=(2\, \unit{\per\second}, 4\, \unit{\per\second})^{\top}$.
    {\interfootnotelinepenalty=10000
\footnote{Note that we introduced the birth rate $\phi_2$ as given in units of per second, \ie, $[\phi_2]=\unit{\per \second}$, which is somewhat non-standard, compared to a reparameterized version ${\phi}_2^\prime=\phi_2 / \rho$ often used in the literature \citep{anderson2015stochastic}, which is of units per molecule second, \ie, $[{\phi}_2^\prime]=\unit{\per \second \per \molec}$.}
}
Therefore, the \ac{hme} in \cref{eq:hme_general} computes to
\begin{equation*}
    \begin{multlined}
        \partial_t \prob (u, v, t \mid \mathcal{H})=\phi_1 \frac{X(0)+\eta u- \rho v}{\rho}\\
        \cdot \left(\frac{1}{2} \partial_v^2  \prob (u, v, t \mid \mathcal{H} )-\partial_v  \prob (u, v, t \mid \mathcal{H}) \right)\\
        -\phi_1 (\partial_v \prob (u, v, t \mid \mathcal{H})  -\prob (u, v, t \mid \mathcal{H}) ) \\
        +\phi_2 (\prob (u-1, v, t \mid \mathcal{H})- \prob (u, v, t \mid \mathcal{H})),
    \end{multlined}
\end{equation*}
see \cref{sec:app_experiments}.
Further, we assume a Gaussian observation model for the state as
\begin{equation}
    Y_n \mid X_n \sim \NDis (y_n \mid x_n, \sigma^2),
    \label{eq:observation_model_exp}
\end{equation}
where we set the standard deviation to $\sigma=4\, \unit{\molec}$.
The state $X(t)$ is observed at $K=50$ time points $\{t_n\}_{n=1}^K$, which are uniformly distributed in the time interval $[0, T]$, with $T=10\, \unit{\second}$.
The resulting latent ground-truth trajectories and the observations are depicted in \cref{fig:gt_state_traj}.
For the numerical simulation of the diffusion process $V(t)$, we use throughout this paper, if not stated otherwise, a stochastic Runge-Kutta method \citep{rossler2010runge}, where we set the integration step to $10^{-2}\, \unit{\second}$, utilizing torchsde\footnote{\url{https://github.com/google-research/torchsde}} \citep{li2020scalable} within the PyTorch framework \citep{paszke2019pytorch}.
For the \ac{mjp} $U(t)$, we utilize the Doob-Gillespie algorithm \citep{doob1945markoff}, see, \eg, \citet{wil:06}.

\subsection*{Posterior Inference}
We sample from the posterior
\begin{equation*}
    X_{[0, T]}, \Phi \mid Y_{1:\nData} \sim \Prob (\mathrm d x_{[0, T]}, \mathrm d \phi \mid y_{1:\nData}),
\end{equation*}
by using the proposed blocked Gibbs particle smoother, as in \cref{alg:mcmc_sampler}.
We perform $1300$ iterations, where we discard the first $300$ samples to adjust for burn-in of the sampler, yielding $L=1000$ posterior samples.
In each step of the sampler, we run the \ac{sir} particle smoothing step with $M=5000$ particles.
To adjust for particle degeneracy, we use systematic resampling inside the particle filtering step and use a minimum effective particle ratio of $\alpha=0.5$ for the resampling threshold, for details see \cref{sec:bootstrap_appendix}.
We choose an independent prior distribution for the rate parameters $\{\Phi_i\}_{i=1}^2$, which is parameterized as
\begin{equation*}
    \prob (\phi_1, \phi_2) = \prod_{i=1}^2 \GamDis(\phi_i \mid a, b).
\end{equation*}
We choose a vague prior distribution by specifying a small shape and rate hyper-parameter, \ie, we use $a=10^{-6}$ and $b=10^{-6}\, \unit{\second}$, respectively.
This yields a vague scale prior that is approximately a flat improper prior distribution, as the gamma distribution with small shape and rate parameters is roughly the \emph{reciprocal distribution} (or log-uniform distribution) on the positive reals, \ie,
\begin{equation*}
    \begin{aligned}
        &\prod_{i=1}^2 \GamDis(\phi_i \mid 10^{-6}, 10^{-6}) \\
        &\approx \prod_{i=1}^2 \LUDis(\phi_i) {\propto} (\phi_1 \phi_2)^{-1}.
    \end{aligned}
\end{equation*}
Therefore, we have a sensible prior that is an improper uniform prior on the real numbers in the log-domain, \ie,
\begin{equation*}
    \begin{aligned}
        &\prob (\log \phi_i) \approx \UniformDis(\log \phi_i)  \propto 1 &&i=1, 2.
    \end{aligned}
\end{equation*}
For sampling from the unnormalized full-conditionals of the parameters in \cref{eq:full_conditional_slow,eq:full_conditional_fast}, we use the \acf{nuts} of \citet{hoffman2014no}, by implementing our system in the probabilistic programming language Pyro \citep{bingham2019pyro}.
The hyper-parameters are set to the default values in Pyro.
In each Gibbs step over the parameters, we perform $100$ warmup steps within \ac{nuts} for burn-in.
In the model for the unnormalized full-conditional of the fast reaction, see \cref{eq:full_conditional_fast}, we compute the reparameterization in \cref{eq:repara_chib} using the step-size of $10^{-2}\, \unit{\second}$,  that is consistent to the particle simulation step size of the stochastic Runge-Kutta integrator.
Subsequently, using the Euler-Maruyama method with the same step size, we integrate the resulting reparameterization in \cref{eq:v_reparam_conditional_fast}.

\subsection{Results}
\begin{figure}
    \centering
    \includegraphics[trim=.2cm 0 .2cm 0,clip]{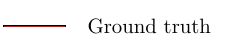}\hfill
    \includegraphics[trim=.2cm 0 .2cm 0,clip]{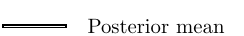}
    \includegraphics{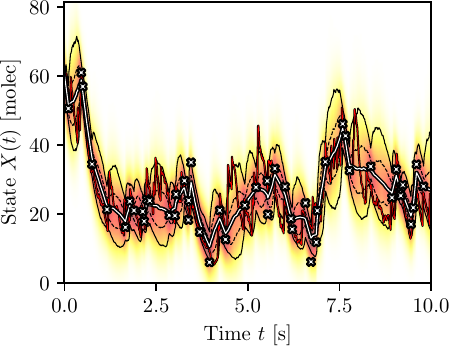}
    \caption{Posterior state inference for the multi-scale birth-death process.
    The plots visualize the ground truth state trajectory, the observations (white crosses), the posterior mean, and the marginal state posterior $\prob(x,t \mid y_{1:\nData})$.
    The figure background color indicates a kernel density estimate for the marginal state posterior and the $5\%\backslash95\%$ (solid line) and $25\%\backslash75\%$ (dashed line) posterior quantiles.}\label{fig:state_inference}
\end{figure}

\begin{figure}
    \centering
    \includegraphics[trim=.2cm 0 .2cm 0,clip]{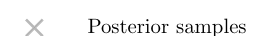}\hfill
    \includegraphics[trim=.2cm 0 .2cm 0,clip]{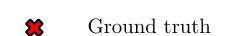}
    \includegraphics{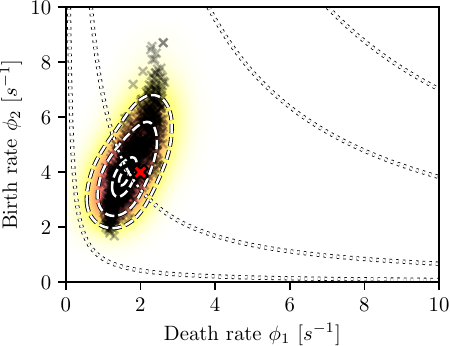}
    \caption{Parameter inference for the multi-scale birth-death process, with only $\nData=50$ observation points.
    The graphic shows the ground-truth parameter and the marginal parameter posterior $\prob(\phi_1,\phi_2 \mid y_{1:\nData})$.
    The marginal parameter posterior is visualized by the posterior parameter samples, and a kernel density estimate is shown in the background.
    The isolines visualize high-density regions for the parameter prior distribution $\prob(\phi_1,\phi_2)\propto (\phi_1 \phi_2)^{-1}$ (dotted white line) and the marginal parameter posterior (dashed white line).}\label{fig:parameter_inference}
\end{figure}

\begin{figure}
    \centering
    \includegraphics{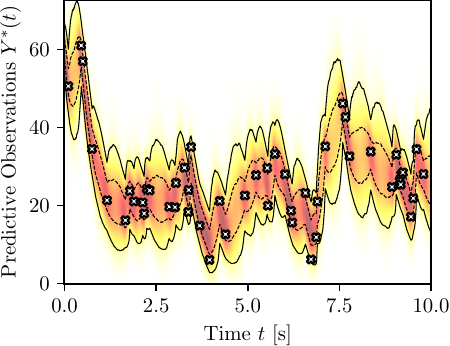}
    \caption{Posterior predictive distribution for the multi-scale birth-death process. The background shows the posterior predictive distribution $\prob(y^\ast, t \mid y_{1:\nData})$ obtained using the posterior samples of the Gibbs sampling procedure. The lines indicate the $5\%\backslash95\%$ (solid line) and $25\%\backslash75\%$ (dashed line) posterior quantiles. The given observations $Y_{1:\nData}$ are shown as white crosses.}\label{fig:predictive_inference}
\end{figure}

The results for the inference of the partially observed multi-scale birth death reaction network using the posterior samples $\{X^{(m)}_{[0,T]},\Phi^{(m)}\}_{m=1}^L$ are depicted in \cref{fig:state_inference,fig:parameter_inference,fig:predictive_inference}.

In \cref{fig:state_inference}, we show the ground-truth latent state trajectory, together with the observations.
The posterior distribution is summarized in the graphic by the posterior mean estimate
\begin{equation*}
    \hat{X}(t)=\Eof*{X(t) | Y_{1:\nData}}\approx \frac{1}{L} \sum_{m=1}^L X^{(m)}(t)
\end{equation*}
and the time-point-wise posterior state marginals
\begin{equation*}
    \prob(x,t \mid y_{1:\nData})\approx \frac{1}{L}\sum_{m=1}^{L} \delta(X^{(m)}(t) - x).
\end{equation*}
We visualize these in \cref{fig:state_inference} by the $5\%\backslash95\%-$ and $25\%\backslash75\%-$quantile regions and by a kernel density approximation $ \hat{\prob}(x,t)  \approx \prob(x,t \mid y_{1:\nData})$, \ie,
\begin{equation*}
    \hat{\prob}(x,t) = \frac{1}{L}\sum_{m=1}^L \mathcal K(X^{(m)}(t) - x),
\end{equation*}
using a scaled Gaussian kernel $\mathcal K(x)=\NDis(x \mid 0, h^2)$, with bandwidth $h$.

In both plots, we observe that the state posterior tracks the ground truth, while the posterior uncertainty increases between observation time points.
Note that due to the observation variance $\sigma^2$, the posterior variance never shrinks exactly to zero.

In \cref{fig:parameter_inference}, we visualize the results for the parameter estimation by the marginal posterior.
\begin{equation*}
    \prob(\phi_1,\phi_2 \mid y_{1:\nData}) \approx \frac{1}{L}\sum_{m=1}^{L} \delta(\Phi_1^{(m)}-\phi_1) \delta(\Phi_2^{(m)}-\phi_2).
\end{equation*}
We show the ground-truth parameters together with the posterior samples $\{\Phi^{(m)}\}_{m=1}^L$.
The marginal parameter posterior is visualized by a kernel density estimate $\hat{\prob}(\phi_1,\phi_2) \approx \prob (\phi_1,\phi_2\mid y_{1:\nData})$, \ie,
\begin{equation*}
    \hat{\prob}(\phi_1,\phi_2)=\hat{\prob}(\phi) = \frac{1}{L}\sum_{m=1}^L \mathcal K(\Phi^{(m)} - \phi).
\end{equation*}
Additionally, we show high-density regions for both the prior $\prob(\phi_1,\phi_2)$ and marginal parameter posterior distribution $\prob(\phi_1,\phi_2 \mid y_{1:\nData})$, depicted using the isolines of the $5\%-$, $25\%-$ $75\%-$ and $95\%-$quantiles.

We see that the posterior concentrates around the ground-truth value.
Consequently, the isolines are shifting from the prior to the posterior density.
However, the parameters cannot be identified due to the limited number of observations $K$ and the observation variance $\sigma^2$.
As such, the parameter posterior samples lie on an ellipse, visualized by the kernel density estimate.
This is a known effect in the context of parameter inference in chemical reaction networks, see, \eg, \citet{wil:06}.

Finally, \cref{fig:predictive_inference} shows the observations $Y_{1:\nData}$ and the posterior predictive distribution
\begin{equation*}
    \begin{aligned}
        \prob (y^\ast, t \mid y_{1:\nData})&=\int \prob (y^\ast \mid x) \prob (x, t  \mid y_{1:\nData})\, \mathrm d x \\
        &=\Eof*{\prob (y^\ast \mid X(t)) | Y_{1:\nData}}\\
        &\approx \frac{1}{L}\sum_{m=1}^{L} \prob (y^\ast \mid X^{(m)}(t) )\\
        &\quad= \frac{1}{L}\sum_{m=1}^{L} \NDis(y^\ast \mid X^{(m)}(t), \sigma^2).
    \end{aligned}
\end{equation*}

From \cref{fig:predictive_inference} we assert that the posterior predictive distribution
\begin{equation*}
    Y^\ast(t) \mid Y_{1:\nData} \sim  \prob (y^\ast, t \mid y_{1:\nData})
\end{equation*}
over hypothetical observations $Y^\ast(t)$ can explain the given observations $Y_{1:\nData}$.

Additionally, to the presented setting, we provide a comparison for different number of observations $K$ and different observation noise parameters $\sigma$ in \cref{table1,table2,table3}.

In \cref{table1}, we give the means of the posterior samples $\{\Phi^{(m)}\}_{m=1}^L$ together with the posterior standard deviations for the ground-truth setting with parameters $(\phi_1,\phi_2)=(2,4)^{\top}$ in terms of  the different number of observations $K$ and the different observation noise standard deviation $\sigma$. 
The results show that the performance of the algorithm increases proportionally with the increase of the number  of observations $K$  and the decrease of the observation noise standard deviation 
$\sigma$. For fixed values of $\sigma$, the increase in the number of observations $K$ gives better results.
It must also be noted that for a fixed number of observations $K$, the increase in the observation noise standard deviation $\sigma$ leads to wider ranges for the posterior mean that brings along uncertainty. 
However, for very large noise and a low number of observations, the parameters get more and more unidentifiable.

\begin{table*}
  \centering
  \begin{tabular}{@{}lrrr@{}}
  \toprule
    & \multicolumn{3}{c}{Posterior parameter mean $\pm$ standard deviation ($\unit{\per\second}$)}\\
     \cmidrule{2-4}
    & \multicolumn{1}{c}{ $\nData=100$} & \multicolumn{1}{c}{ $\nData=50$} & \multicolumn{1}{c}{ $\nData=10$} \\
    \cmidrule{2-4}
    $\sigma=1\, \unit{\molec}$ & $(1.90,4.44) \pm (0.10,0.72)$ & $(1.70,4.27) \pm (0.12,0.77)$ & $(0.59,1.39) \pm (0.13,0.55)$ \\
    $\sigma=2\, \unit{\molec}$  & $(2.16,4.02) \pm (0.30,0.85)$ & $(2.78,5.66) \pm (0.36, 1.13)$ & $(2.08, 3.72) \pm (0.50,1.26)$  \\
    $\sigma=4\, \unit{\molec}$ & $(1.06,2.77) \pm (0.11,0.66)$ & $(1.55,2.24) \pm (0.37,0.78)$ & $(2.69,5.78) \pm (0.84,2.20)$  \\
    $\sigma=8\, \unit{\molec}$ & $(2.12,3.93) \pm (0.52,1.22)$ & $(1.63,4.45) \pm (0.45,1.52)$ & $(0.20,0.05) \pm (0.06,0.16)$ \\
    \bottomrule
  \end{tabular}
  \caption{Posterior means and ($\pm$) standard deviations  for the parameters $(\phi_1,\phi_2)=(2,4)^{\top}$, depending on the number of observations $K$ and the observation noise standard deviation $\sigma$.}\label{table1}
\end{table*}

In \cref{table2}, we compare the mean of the effective
number of particles $\bar M_{\text{EPS}}$ and the mean of the unique number of particles $\bar M_{\text{unique}}$ after resampling with $M=5000$ particles for different
number of observations $K$ and for different observation noise standard deviations $\sigma$. 
The outcomes validate that enough particles always survive after resampling. 
It is visible that $\bar M_{\text{EPS}}$ decreases with the increase of the observation noise standard deviation $\sigma$ and the increase in the number of observation $K$. 
This is caused by the increase in variance of the system dynamics.
Another result  that can be seen from the table is that for a fixed number of observations $K$, the mean of the unique particles $\bar M_{\text{unique}}$ increase parallel with the increase in the observation noise standard deviation  $\sigma$.
While for a fixed observation noise standard deviation $\sigma$,  the mean of the unique particles $\bar M_{\text{unique}}$ increase  with the number of observations $K$.
\begin{table*}
  \centering
  \begin{tabular}{@{}lrrcrrcrr@{}}
  \toprule
    & \multicolumn{2}{c}{ $\nData=100$} & \phantom{abc} & \multicolumn{2}{c}{ $\nData=50$} & \phantom{abc} & \multicolumn{2}{c}{ $\nData=10$} \\
    \cmidrule{2-3}  \cmidrule{5-6} \cmidrule{8-9}
    & $\bar M_{\text{EPS}}$ & $\bar M_{\text{unique}}$ && $\bar M_{\text{EPS}}$ & $\bar M_{\text{unique}}$&& $\bar M_{\text{EPS}}$ & $\bar M_{\text{unique}}$\\
    \cmidrule{2-9}
    $\sigma=1\, \unit{\molec}$ & $4644$ & $2138$  && $4806$ & $1662$ &&  $4988$ & $1542$ \\
    $\sigma=2\, \unit{\molec}$ & $4389$ & $2960$ && $4704$ & $2224$ && $4934$ & $1887$    \\
    $\sigma=4\, \unit{\molec}$  & $4199$ & $3612$  && $4251$ & $3494$ && $4744$ & $2524$  \\
    $\sigma=8\, \unit{\molec}$  & $4057$ & $4164$  && $4072$ & $3918$ && $4087$ & $4569$  \\
    \bottomrule
  \end{tabular}
  \caption{The Mean of the effective number of particles $\bar M_{\text{EPS}}$ and the mean of the unique number of particles $\bar M_{\text{unique}}$ after resampling used for the particle filtering with $M=5000$ particles, depending on the number of observations $K$ and the observation noise standard deviation $\sigma$.} \label{table2}
\end{table*}

Finally, in \cref{table3}, we compare the root mean square error $\text{RMSE}=\sqrt{\frac{1}{T}\int_{0}^T (\hat{X}(t)-X(t))^2\, \mathrm d t}$  of the state estimate $ \hat{X}(t)$ for different number of observations $K$ and for different observation noise standard deviations $\sigma$.
It is discernible that for fixed values of the observation noise standard deviation $\sigma$, $\text{RMSE}$ decreases with the increase in the number of observation $K$. 
Also, the decrease in the observation noise standard deviation  $\sigma$ for a fixed number of observations $K$ results in a decrease in the $\text{RMSE}$.
Note, that the presented  $\text{RMSE}$ value, is only given for one experiment.
\begin{table}
  \centering
  \begin{tabular}{@{}lrrr@{}}
  \toprule
       &\multicolumn{3}{c}{ $\text{RMSE}$ ($\unit{\molec}$)}\\
    \cmidrule{2-4}
 & \multicolumn{1}{c}{ $\nData=100$} & \multicolumn{1}{c}{ $\nData=50$} & \multicolumn{1}{c}{ $\nData=10$} \\
    \cmidrule{2-4}
    $\sigma=1\, \unit{\molec}$ & $3.84$ & $4.91$ & $9.69$ \\
    $\sigma=2\, \unit{\molec}$  & $3.77$ & $4.59$ & $5.87$ \\
    $\sigma=4\, \unit{\molec}$ & $5.55$ & $4.12$ & $10.58$ \\
    $\sigma=8\, \unit{\molec}$  & $5.47$ & $6.62$ & $9.06$ \\
    \bottomrule
  \end{tabular}
  \caption{Root mean square error $\text{RMSE}=\sqrt{\frac{1}{T}\int_{0}^T (\hat{X}(t)-X(t))^2\, \mathrm d t}$  of the posterior mean state estimate $\hat X(t)$, depending on the number of observations $K$ and the observation noise standard deviation $\sigma$.} \label{table3}
\end{table}
    \section{Conclusion}
\label{sec:con}
By exploiting a hybrid modeling approach to \acp{brn} we presented a coherent framework for fast and tractable Bayesian inference for partially observed reaction networks exhibiting a multi-scale behavior.
The proposed blocked Gibbs particle smoothing algorithm overcomes the obstacles posed by the derived intractable equations of exact posterior inference.
This is achieved by performing separate blocked Gibbs steps for state and parameter inference in the \acp{brn} modeled by a jump-diffusion approximation.
Efficient inference is accomplished by utilizing a particle-based forward-filtering backward-smoothing algorithm and \iac{mcmc}-based sampler for state and parameter inference, respectively.
The presented numerical case study exemplifies the algorithm by showing its applicability to an illustrative setup of a birth-death process, which exhibits a multi-scale behavior.

As a possible future work, we think that our algorithm can be the base for new inference algorithms exploiting the jump-diffusion approximation for \ac{brn} models.
For example, it is known that a naive application of a particle-based posterior approximation suffers as the state dimension increases.
Therefore, in order take to make the proposed algorithm more applicable in such settings, a possible future work includes the improvement of the state inference procedure, \eg, by finding an improved proposal distribution for the underlying particle approximation.
Additionally, we think that the ideas presented in this work might be of use in other contexts for state and parameter inference, where the underlying model exhibits a multi-scale behavior, well beyond \acp{brn}.

     \backmatter
     \bmhead{Acknowledgments} 
     Derya Alt{\i}ntan  acknowledges support from the German Academic Exchange Service DAAD, Program no: 57552334 Grant no.\ 91527068. 
     Additionally, this work has been funded by the German Research Foundation (DFG) as part of the project B4 within the Collaborative Research Center (CRC) 1053 – MAKI.
    \begin{appendices}
        \section{Notation}
\label{notation}
In this section, we present some basic notation used throughout this paper.

All random variables and their realizations are represented by upper-case symbols, \eg, $Z$, and lower-case symbols, \eg, $z$, respectively.

We denote by $\unitvector_j$ and $\bar{\unitvector}_j$ the $\nSlow$-dimensional and the $(\nReactions-\nSlow)$-dimensional unit vectors with $1$ in the $j$th component and $0$ in all other components, respectively.

Any sequence $z_{a:b}$, with $a\in \mathbb N$, $b\in \mathbb N$ and $a<b$ represents the vector
\begin{equation*}
    z_{a:b}\coloneqq(z_a, z_{a+1}, \dots, z_{b-1}, z_b)^\top.
\end{equation*}
For a stochastic process $Z$, we write $Z_{[a, b]}$ representing the path
\begin{equation*}
    Z_{[a, b]} \coloneqq \{Z(t):\, t \in [a, b]\}.
\end{equation*}
We denote by
\begin{equation*}
    Z_i=Z(t_i),
\end{equation*}
with $i \in \mathbb N$ a random variable $Z_i$ at an observation point $t_i$ of a stochastic process $Z$, with the observation times $t_1 <t_2<t_3< \dots$ and we define $t_0 \coloneqq 0$.

For the probability measure of a random variable $Z$, we use the shorthand %
\begin{equation*}
    \begin{aligned}
        \Prob (\mathrm d z) \coloneqq \Prob (Z \in \mathrm d z)
    \end{aligned}
\end{equation*}
that is for a set $\mathcal A$ we have $\Prob (Z \in \mathcal A)= \int_{\mathcal A} \Prob (\mathrm d z)$.
For the probability mass function of a time-dependent discrete-valued variable $U(t)$ at time $t$ we write
\begin{equation*}
    \prob (u, t)\coloneqq \Prob (U(t)=u).
\end{equation*}
For the probability density function of a time-dependent continuous-valued random variable $V(t)$ at time $t$, we use
\begin{equation*}
    \prob (v, t) \coloneqq \partial_v \Prob (V(t) \leq v).
\end{equation*}
where \enquote{$\leq$} and \enquote{$\partial_v$} denote element-wise operation.
For the joint probability density function with a discrete random variable $U(t)$ and a continuous random variable $V(t)$ at time $t$, we use
\begin{equation*}
    \prob (u, v, t) \coloneqq \partial_v \Prob (V(t) \leq v, U(t)=u).
\end{equation*}
with $p: \mathbb{N}^{\nSlow}\times \mathbb{R}^{(A-\nSlow)}_{\geq 0} \longrightarrow \mathbb{R}_{\geq 0} $.
For the joint distribution, we have
\begin{equation*}
    \prob (u, v, t)=\prob (v \mid u, t)\prob (u, t),
\end{equation*}
where the conditional density is given as %
\begin{equation*}
    \prob (v \mid u, t) \coloneqq \partial_v \Prob (V(t) \leq v \mid U(t)=u).
\end{equation*}
For the conditional density at two different time points, we use
\begin{equation*}
    \begin{aligned}
        &\prob (u, v, t \mid u', v', t') \coloneqq \\
        &\partial_v \Prob (V(t) \leq v, U(t)=u \mid V(t')=v', U(t')=u').
    \end{aligned}
\end{equation*}

\section{Reaction Network Model}
\label{sec:reaction_model_appendix}

\subsection{Proof of \texorpdfstring{\cref{thm_hybrid}}{Theorem 1}}

\begin{proof}
    \label{sec:proof_hme}
    Following \citet{ak:20} and \citet{paw:67}, we write
    \begin{equation*}
        \begin{aligned}
            &\partial_t \prob (u, v, t\mid \mathcal{H})\\
            &= \sum_{u' \in \mathcal{U}} \lim_{h \rightarrow 0} \frac{1}{h} [ \prob  (u, t+h \mid u', v, t, \mathcal{H})- \delta_{uu'}]\\
            &\; \cdot \prob (u', v, t\mid \mathcal{H}) +\sum_{a_{1}, a_{2}, \dots, a_{\nReactions-\nSlow}=1}^{ \infty} \left( \prod_{i=1}^{\nReactions-\nSlow} \frac{(-1)^{a_{i}} \frac{\partial^{a_{i}}}{\partial v_{i}^{a_{i}}}}{a_{i}!}\right)  \\
            &\quad
            \begin{multlined}
                \cdot \lim_{h \rightarrow 0} \frac{1}{h} \EofLeft*{\prod_{i=1}^{\nReactions-\nSlow} \{V_{i}(t+h)-V_{i}(t)\}^{a_{i}} | U(t)=u,} \\ \EofRight*{V(t) \leq v, U(t+h)=u, \mathcal{H} },
            \end{multlined}
        \end{aligned}
    \end{equation*}
    with $\Eof*{V(t) | u}={\int_{ \mathbb{R}_{ \geq 0}^{\nReactions-\nSlow}}} v \: \prob (v \mid u, t, \mathcal{H} ) \, \mathrm dv$.

    It is proven in \citet{paw:67}, that for $ \sum_{i=1}^{\nReactions-\nSlow} a_{i} \geq 3$ we have
    \begin{equation*}
        \begin{multlined}
            \lim_{h \rightarrow 0} \frac{1}{h} \EofLeft*{\prod_{i=1}^{\nReactions-\nSlow} \{V_{i}(t+h)-V_{i}(t)\}^{a_{i}} | U(t)=u,}\\
            \EofRight*{V(t) \leq v, U(t+h)=u, \mathcal{H} }=0.
        \end{multlined}
    \end{equation*}
    This result gives us
    \begin{equation*}
        \begin{aligned}
            &\partial_t \prob (u, v, t\mid \mathcal{H})\\
            &{}= \sum_{u' \in \mathcal{U}} \lim_{h \rightarrow 0} \frac{1}{h} [ \prob  (u, t+h \mid u', v, t, \mathcal{H})- \delta_{uu'}]\\
            &\quad \cdot\prob (u', v, t\mid \mathcal{H})- \sum_{j=1}^{\nReactions-\nSlow} \partial_{v_{j}} [\Lambda_{j} \prob (u, v, t \mid \mathcal{H})]\\
            &\quad+ \frac{1}{2} \sum_{i, j=1}^{\nReactions-\nSlow} \partial_{v_{i}} \partial_{v_{j}} [\Lambda_{ij} \prob (u, v, t \mid \mathcal{H})],
        \end{aligned}
    \end{equation*}
    where

    \begin{equation*}
        \begin{aligned}
            &\begin{multlined}
                 \Lambda_{j}=\lim_{h \rightarrow 0} \frac{1}{h} \EofLeft*{(V_{j}(t+h)- V_{j}(t)) |
                 U(t)=u,} \\ \EofRight*{V(t) \leq v, U(t+h )=u, \mathcal{H}}
            \end{multlined}\\
            &\begin{multlined}
                 \Lambda_{ij}= \lim_{h \rightarrow 0} \frac{1}{h} \EofLeft*{\{V_{i}(t+h)- V_{i}(t)\} \{V_{j}(t+h)}\\
                 \EofRight*{- V_{j}(t)\} |  U(t)=u, V(t) \leq v, U(t+h )=u, \mathcal{H}}.
            \end{multlined}
        \end{aligned}
    \end{equation*}
    In the presented jump-diffusion approximation, fast reactions always fire. This means that between two successive firing times $\tau_1$ and $\tau_2$ of slow reactions, the reaction counter of a fast reaction satisfies the following diffusion process
    \begin{equation}
        \begin{aligned}
            \label{eq:reaction_counter_dif}
            V(t)&=V(\tau_1)+ \sum_{j \in \mathcal{C}}\left( \int_{\tau_1}^{t} \kappa_{j}(U(s), V(s)) \, \mathrm ds\right)\bar{\unitvector}_{j} \\
            &+ \sum_{j \in \mathcal{C}} W_j \left( \int_{\tau_1}^{t} \kappa_{j}(U(s), V(s))\, \mathrm ds\right) \bar{\unitvector}_{j}.
        \end{aligned}
    \end{equation}
    Based on these results, we obtain $\Lambda_{j}$, $\Lambda_{ij}$ as follows \citep{gill:80, kam:82}
    \begin{equation*}
        \Lambda_{j}= \sum_{k \in \mathcal{C}} \bar{\unitvector}_{jk} \kappa_k(u, v), \quad \Lambda_{ij}= \sum_{k \in \mathcal{C}} \bar{\unitvector}_{ik} \bar{\unitvector}_{jk} \kappa_k(u, v),
    \end{equation*}
    which in turn gives
    \begin{equation}
        \begin{aligned}
            &\partial_t \prob (u, v, t\mid \mathcal{H})\\
            &= \sum_{u' \in \mathcal{U}} \lim_{h \rightarrow 0} \frac{1}{h} [ \prob  (u, t+h \mid u', v, t, \mathcal{H})- \delta_{uu'}] \\ &{}\quad\cdot \prob (u', v, t\mid \mathcal{H}) - \sum_{j \in \mathcal{C}} \partial_{v_j} ( \kappa_{j}(u, v) \prob (u, v, t \mid \mathcal{H}) ) \\
            &\quad +\frac{1}{2} \sum_{j \in \mathcal{C}} \partial_{v_j}^{2} ( \kappa_{j}(u, v) \prob (u, v, t \mid \mathcal{H})).
        \end{aligned}\label{eq:hme1_1}
    \end{equation}
    Now, let us focus on the first summation on the right-hand side of the \cref{eq:hme1_1}. Using $\delta_{uu}=1$, gives the following equality
    \begin{equation*}
        \begin{aligned}
            &\sum_{u' \in \mathcal{U}} \lim_{h \rightarrow 0} \frac{1}{h} [ \prob  (u, t+h \mid u', v, t, \mathcal{H})- \delta_{uu'}]\\
            &=\sum_{\substack{
                u \neq u' \\
                u' \in \mathcal{U}}} \lim_{h \rightarrow 0} \frac{1}{h} [ \prob  (u, t+h \mid u', v, t, \mathcal{H})]\\
            &+ \lim_{h \rightarrow 0} \frac{1}{h} [ \prob  (u, t+h \mid u, v, t, \mathcal{H})-1].
        \end{aligned}
    \end{equation*}

    Further, by exploiting the complement rule we write
    \begin{equation*}
        \begin{aligned}
            &\lim_{h \rightarrow 0} \frac{1}{h} [ \prob  (u, t+h \mid u, v, t, \mathcal{H})-1]\\
            &= \lim_{h \rightarrow 0} \frac{1}{h}
            [ - \sum_{\substack{
                u \neq u' \\
                u' \in \mathcal{U}}} \prob (u', t+h\mid u, v, t, \mathcal{H}) ].
        \end{aligned}
    \end{equation*}
    Then, we get
    \begin{equation*}
        \begin{aligned}
            &\sum_{u' \in \mathcal{U}} \lim_{h \rightarrow 0} \frac{1}{h} [ \prob  (u, t+h \mid u', v, t, \mathcal{H})- \delta_{uu'}]\\
            &= \sum_{\substack{
                u \neq u' \\
                u' \in \mathcal{U}}} \Big( \lim_{h \rightarrow 0} \frac{1}{h} [ \prob  (u, t+h \mid u', v, t, \mathcal{H})] \\
            & -\lim_{h \rightarrow 0} \frac{1}{h} [ \prob (u', t+h, \mid u, v, t, \mathcal{H}) ] \Big).
        \end{aligned}
    \end{equation*}
    If $U(t)=u$, then one firing of reaction $R_i$, $i \in \mathcal{D}$, will jump to the state $u'=u+\unitvector_i$ which gives
    \begin{equation*}
        \lim_{h \rightarrow 0} \frac{1}{h} [ \prob (u', t+h, \mid u, v, t, \mathcal{H}) ]= \kappa_i(u, v),
    \end{equation*}
    and similarly for $u'=u-\unitvector_i$
    \begin{equation*}
        \lim_{h \rightarrow 0} \frac{1}{h} [ \prob (u, t+h, \mid u', v, t, \mathcal{H}) ]= \kappa_i(u-\unitvector_i, v).
    \end{equation*}
    Substitution these results into \cref{eq:hme1_1} give us
    \begin{equation*}
        \partial_t \prob (u, v, t \mid \mathcal{H}) =\mathscr{A} \prob (u, v, t \mid \mathcal{H}),
    \end{equation*}
    where $\mathscr{A}(\cdot)=\mathscr{D} (\cdot)+\mathscr{C} (\cdot)$ defined as
    \begin{equation*}
        \begin{aligned}
            \mathscr{D} \prob (u, v, t \mid \mathcal{H}) &= \sum_{i \in \mathcal{D}} \left(\kappa_{i}(u-\unitvector_i, v) \prob (u-\unitvector_i, v, t \mid \mathcal{H}) \right.\\
            &\left. -\kappa_{i}(u, v)\prob (u, v, t \mid \mathcal{H}) \right) \\
            \mathscr{C} \prob (u, v, t \mid \mathcal{H}) &= -\sum_{j \in \mathcal{C}} \partial_{v_j} ( \kappa_{j}(u, v) \prob (u, v, t \mid \mathcal{H})) \\
            &+\frac{1}{2} \sum_{j \in \mathcal{C}} \partial_{v_j}^{2} ( \kappa_{j}(u, v) \prob (u, v, t \mid \mathcal{H}))
        \end{aligned}
    \end{equation*}
    which completes the proof.
\end{proof}

\subsection{Computing the Radon-Nikodym Derivative \texorpdfstring{$D(u_{[0, T]})$}{}}
\label{sec:radon_nikodym}
Here, we present the derivation of the Radon-Nikodym derivative
\begin{equation*}
    D(u_{[0, T]})= \frac{\mathrm d \Prob_{U\mid V,\Phi}}{\mathrm d \Prob_{\zeta}},
\end{equation*}
between the path measures $\Prob_{U\mid V,\Phi}$ and $\Prob_{\zeta}$, of the counting process $U \mid V,\Phi$ and the unit Poisson process $\zeta$, respectively.
For this, we divide the interval $[0, T]$ into sub-intervals $[t_k, t_{k+1}]$, with $t_k=k \Delta t$, $k=0, 1, \dots, a$.
Next, we obtain a discrete-time approximation for $D(u_{[0, T]})$ as
\begin{equation*}
    D_{\Delta t}(u_{0:a})= \frac{{\prob}_{U \mid V, \Phi}(u_{0:a})}{{\prob}_{\zeta}(u_{0:a})},
\end{equation*}
for which taking the continuous-time limit yields the thought after density expression, \ie,
\begin{equation*}
    \lim_{\Delta t \rightarrow 0} D_{\Delta t}(u_{0:a}) = D(u_{[0, T]}).
\end{equation*}
In the following, we use the notation $Z_k=Z(t_k)$ with components $Z_{k, i}=Z_i(t_k)$, $i=1, 2, \dots, \ell$, and $\{Z_j\}_{j=0}^{a}=\{Z_1, Z_2, \dots, Z_a\}$
for any process $Z$. Finally, for the conditional probability of any discrete process $Z$, we use the shorthand
\begin{equation*}
    \begin{aligned}
        & \begin{multlined}
              \Prob (Z_k-Z_{k-1}=\Delta z \mid z_{k-1})
        \end{multlined}\\
        &\begin{multlined}
             \coloneqq \Prob (Z_k-Z_{k-1}= \Delta z   \mid \{V_j=v_j\}_{j=0}^{k},\\
             Z_{k-1}=z_{k-1})
        \end{multlined}
    \end{aligned}
\end{equation*}
where $\Delta z$ is an $\nSlow$-dimensional vector.

Based on \citet{ak:11}, we obtain the following results for the reaction counting processes in a small time interval $[t, t+\Delta t)$.
For the process $U$, we have the following expressions
\begin{align*}
    \begin{aligned}
        &\begin{multlined}
             \Prob (U_k-U_{k-1}=e_i \mid u_{k-1}) \approx \kappa_{i}(u_{k-1}, v_{k-1}) \Delta t
        \end{multlined} \\
        &\begin{multlined}
             \Prob (U_k-U_{k-1}=0 \mid u_{k-1}) \\
             \approx  \exp (- \sum_{i \in \mathcal{D}} \kappa_{i}(u_{k-1},v_{k-1}) \Delta t ).
        \end{multlined}
    \end{aligned}
\end{align*}
Similarly, we obtain for the stochastic process $\zeta$
\begin{align*}
    \begin{aligned}
        &\Prob (\zeta_k-\zeta_{k-1}=e_i \mid u_{k-1}) \approx \Delta t \\
        &\Prob (\zeta_k-\zeta_{k-1}=0 \mid u_{k-1}) \approx \exp (- \sum_{i \in \mathcal{D}} \Delta t ).
    \end{aligned}
\end{align*}
This gives us the probability distribution for $\{U_k\}_{k=0}^a$ over the grid as
\begin{equation*}
    \begin{aligned}
        & {\prob}_{{U \mid V, \Phi}}(u_{0:a})= \prob_{U \mid V, \Phi}(u_0, u_1, \dots, u_a \mid \{v_j\}_{j=0}^{a} ) \\
        &= \prob_{U \mid V, \Phi}(u_0) \prod_{k=1}^{a} \prob_{U \mid V, \Phi}(u_k \mid u_{k-1}, \{v_j\}_{j=0}^{a}) \\
        &\approx \delta_{u_{0,0}} \prod_{k=1}^{a}\left\{\delta_{u_{k-1}, u_{k}} \Prob (U_k-U_{k-1}=0 \mid u_{k-1}) \right. \\
        &+ \left. \sum_{i \in \mathcal{D}} \delta_{u_{k-1}+e_i, u_{k}} \Prob (U_k-U_{k-1}=e_i \mid u_{k-1}) \right\} \\
        & \approx \delta_{u_{0, 0}} \prod_{k=1}^{a}
        \left \{\delta_{u_{k-1}, u_{k}} \exp (-\sum_{i \in \mathcal{D}} \kappa_{i}(u_{k-1}, v_{k-1}) \right. \\
        &\left. \cdot \Delta t) + \sum_{i \in \mathcal{D}} \delta_{u_{k-1}+e_i, u_{k}} \kappa_{i}(u_{k-1}, v_{k-1})\Delta t \right \}
    \end{aligned}
\end{equation*}
where $\delta_{u_i, u_j}$ is the Kronecker delta function and $u_0=0$. Similarly, we get the following equation for the distribution of $\{\zeta_k\}_{k=0}^a$ over the grid
\begin{equation*}
    \begin{aligned}
        & {\prob}_{\zeta}(u_{0:a})=\prob_{\zeta}(u_0, u_1, \dots, u_a \mid \{v_j\}_{j=0}^{a} ) \\
        & = \prob_{\zeta}(u_0) \prod_{k=1}^{a} \prob_{\zeta}(u_k \mid u_{k-1}, \{v_j\}_{j=0}^{a}) \\
        &\approx \delta_{u_{0, 0}} \prod_{k=1}^{a} \left \{\delta_{u_{k-1}, u_{k}} \Prob (\zeta_k-\zeta_{k-1}=0 \mid u_{k-1}) \right. \\
        &+ \left. \sum_{i \in \mathcal{D}} \delta_{u_{k-1}+e_i, u_{k}} \Prob (\zeta_k-\zeta_{k-1}=e_i \mid u_{k-1}) \right \} \\
        &\begin{multlined}
             \approx \delta_{u_{0, 0}} \prod_{k=1}^{a} \left \{\delta_{u_{k-1}, u_{k}} \exp (-\sum_{i \in \mathcal{D}} \Delta t )\right. \\ \left. + \sum_{i \in \mathcal{D}} \delta_{u_{k-1}+e_i, u_{k}} \Delta t \right \}
        \end{multlined}
    \end{aligned}
\end{equation*}
Now, we obtain the following discrete-time approximation for the Radon-Nikodym derivative
\begin{equation*}
    \begin{aligned}
        &D_{\Delta t}(u_{0:a})=\frac{{\prob}_{U \mid V, \Phi}(u_{0:a})}{{\prob}_{\zeta}(u_{0:a})} \\
        &\approx \left(\delta_{u_{0, 0}} \prod_{k=1}^{a} \left \{\delta_{u_{k-1}, u_{k}} \exp \Big(-\sum_{i \in \mathcal{D}} \kappa_{i}(u_{k-1},  \right. \right. \\
        &  v_{k-1}) \Delta t \Big) + \left. \left. \sum_{i \in \mathcal{D}} \delta_{u_{k-1}+e_i, u_{k}} \kappa_{i}(u_{k-1}, v_{k-1})\Delta t \right\} \right)\\
        & \cdot \left(\delta_{u_{0, 0}} \prod_{k=1}^{a} \left \{\delta_{u_{k-1}, u_{k}} \exp \left(-\sum_{i \in \mathcal{D}} \Delta t\right) \right. \right. \\
        &+ \left. \left. \sum_{i \in \mathcal{D}} \delta_{u_{k-1}+e_i, u_{k}} \Delta t \right \}\right)^{-1}
    \end{aligned}
\end{equation*}
By using the fact that if $\delta_{u_{k-1}, u_{k}}=0$, then $\delta_{u_{k-1}+e_i, u_{k}}=1$ or if $\delta_{u_{k-1}, u_{k}}=1$, then $\delta_{u_{k-1}+e_i, u_{k}}=0$, we write
\begin{equation*}
    \begin{aligned}
        &D_{\Delta t}(u_{0:a})=\frac{{\prob}_{U \mid V, \Phi}(u_{0:a})}{{\prob}_{\zeta}(u_{0:a})} \\
        &\approx \prod_{k=1}^{a} \left\{\delta_{u_{k-1}, u_{k}} \frac{ \exp \left(-\sum_{i \in \mathcal{D}} \kappa_{i}(u_{k-1}, v_{k-1}) \Delta t\right)}{\exp (-\sum_{i \in \mathcal{D}} \Delta t)} \right. \\
        &+ \left. \sum_{i \in \mathcal{D}} \delta_{u_{k-1}+e_i, u_{k}} \frac{\kappa_{i}(u_{k-1}, v_{k-1}) \Delta t} {\Delta t} \right \} \\
        & \approx \prod_{k=1}^{a} \left \{\delta_{u_{k-1}, u_{k}} \exp ( \sum_{i \in \mathcal{D}} [1- \kappa_{i}(u_{k-1}, v_{k-1})] \Delta t ) \right. \\
        & \left. + \sum_{i \in \mathcal{D}} \delta_{u_{k-1}+e_i, u_{k}} \kappa_{i}(u_{k-1}, v_{k-1})\right \} \\
        & \approx \prod_{k=1}^{a} \left \{ \exp ( \sum_{i \in \mathcal{D}} [1- \kappa_{i}(u_{k-1}, v_{k-1})] \Delta t )^{\delta_{u_{k-1}, u_{k}}} \right. \\
        &\quad \left. \prod_{i \in \mathcal{D}} \kappa_{i}(u_{k-1}, v_{k-1})^{\delta_{u_{k-1}+e_i, u_{k}}}
        \right \}\\
        & \approx \exp \left( \sum_{k=1}^{a} {\delta_{u_{k-1}, u_{k}}} \sum_{i \in \mathcal{D}} [1- \kappa_{i}(u_{k-1}, v_{k-1})] \Delta t \right) \\
        & \cdot \prod_{k=1}^{a} \prod_{i \in \mathcal{D}} \kappa_{i}(u_{k-1}, v_{k-1})^{\delta_{u_{k-1}+e_i, u_{k}}}.
    \end{aligned}
\end{equation*}
By taking the continuous-time limit we obtain a Riemann integral as
\begin{equation*}
    \begin{aligned}
        &\exp \left( \sum_{k=1}^{a} {\delta_{u_{k-1}, u_{k}}} \sum_{i \in \mathcal{D}} [1- \kappa_{i}(u_{k-1}, v_{k-1})] \Delta t \right) \\ &\stackrel{ \Delta t \rightarrow 0}{\longrightarrow} \exp \left(\int_{0}^{T} \sum_{i \in \mathcal{D}} [1-\kappa_i(u(s), v(s))]\, \mathrm d s \right).
    \end{aligned}
\end{equation*}
Note that $u_i(T) \Vert_1$ represent the firing number of the $i$th slow reaction in the time interval $[0, T]$, therefore, we write
\begin{equation*}
    \begin{aligned}
        & \prod_{k=1}^{a} \prod_{i \in \mathcal{D}} \kappa_i(u_{k-1}, v_{k-1})^{\delta_{u_{k-1}+e_i, u_{k}}} \\
        &\stackrel{ \Delta t \rightarrow 0}{\longrightarrow} \prod_{i \in \mathcal D} \prod_{j=1}^{u_i(T)} \kappa_{i}
        (u(\tau_{i,j}^{-}), v(\tau_{i,j}^{-})),
    \end{aligned}
\end{equation*}
where $\tau_{i,j}^{-}$ represents the time right before $\tau_{i,j}$, which is the $j$th firing time of the $i$th slow reaction $R_i$, $i \in \mathcal D$. Hence, for the $j$th firing time $\tau_{i,j}$ of reaction $i$ we have
\begin{equation*}
    \begin{aligned}
        &u(\tau_{i,j})-u(\tau_{i,j}^{-})=e_{i}, &&v(\tau_{i,j})=v(\tau_{i,j}^{-}).
    \end{aligned}
\end{equation*}
Finally, we get
\begin{align*}
    \begin{aligned}
        & D(u_{[0, T]})=\frac{\mathrm d \Prob_{U\mid V,\Phi}}{\mathrm d \Prob_{\zeta}}(u_{[0, T]}) \\
        &= \exp \left(\int_{0}^{T} \sum_{i \in \mathcal{D}} [1-\kappa_i(u(s), v(s))]\, \mathrm d s \right) \\
        &\cdot \prod_{i \in \mathcal D} \prod_{j=1}^{u_i(T)} \kappa_{i}
        (u(\tau_{i,j}^{-}), v(\tau_{i,j}^{-})).
    \end{aligned}
\end{align*}

\section{Posterior Inference}
\label{sec:posterior_inference_appendix}

\subsection{Calculation of the Filtering Distribution}
\label{sec:filts}
We define the filtering distribution as
\begin{equation*}
    \pi(u, v, t) \coloneqq \prob (u, v, t \mid \phi, y_{1:n}),
\end{equation*}
with the density
\begin{equation*}
    \begin{multlined}
        \prob (u, v, t \mid \phi, y_{1:n}) \coloneqq \partial_{v_1} \partial_{v_2} \dots \partial_{v_{\nReactions-\nSlow}} \Prob (V(t) \leq v, \\
        U(t)=u\mid \Phi=\phi, Y_{1:n}=y_{1:n} ),
    \end{multlined}
\end{equation*}
where $n=\max\{n' \in \mathbb{N}\mid t_{n'}\leq t\}$.
Computation of the filtering distribution can be divided into two steps which are the \emph{prediction step} and the \emph{update step}.
The prediction step considers the filtering distribution between the observation time points and the update step at the observation time points.

\subsubsection{The Filtering Distribution Between Observation Points}
In this section, we aim to obtain the filtering distribution in the time interval $[t, t+h]$, $h>0$, without any observation. We have
\begin{equation*}
    \begin{aligned}
        &\begin{multlined}
             \pi(u, v, t+h)
        \end{multlined}\\
        &\begin{multlined}
             = \prob (u, v, t+h\mid \phi, y_{1:n})
        \end{multlined}\\
        &\begin{multlined}
             = \sum_{u' \in \mathcal{U}} \int \prob (u, v, t+h, u', v', t \mid \phi, y_{1:n})\, \mathrm d v'
        \end{multlined}\\
        &\begin{multlined}
             = \sum_{u' \in \mathcal{U}} \int \prob (u, v, t+h \mid u', v', t, \phi, y_{1:n}) \\
             \cdot \prob (u', v', t\mid \phi, y_{1:n}) \, \mathrm d v'.
        \end{multlined}
    \end{aligned}
\end{equation*}
Since we do not have any observations in the interval under consideration, we write
\begin{equation}
    \begin{multlined}
        \pi(u, v, t+h)
        = \sum_{u' \in \mathcal{U}} \int \prob (u, v, t+h \mid u', v', t, \phi)  \\
        \pi(u', v', t)\, \mathrm dv'.
    \end{multlined}
    \label{eq:chapman_filter_app}
\end{equation}
This is the Chapman-Kolmogorov equation, see, \eg, \citet{kah:21}, for the probability distribution $\prob (u, v, t+h \mid u', v', t, \phi)$.
This means that the filtering distribution $\pi(u, v, t)$ between observation points satisfies the \ac{hme}
\begin{equation*}
    \partial_t \pi(u, v, t) =\mathscr{A} \pi(u, v, t).
\end{equation*}

Note that we can specify $t$ and $h$ in \cref{eq:chapman_filter_app} such that we obtain the prediction step
\begin{equation}
    \begin{aligned}
        &\begin{multlined}
             \lim_{t \nearrow t_n} \pi(u,v,t)\coloneqq \pi(u,v,t_n^-)
        \end{multlined}\\
        &\begin{multlined}
             = \prob ( u, v, t_n \mid \phi, y_{1:n-1})
        \end{multlined} \\
        &\begin{multlined}
             = \sum_{u' \in \mathcal{U}} \int_{\mathcal{V}} \prob ( u, v, t_n \mid u', v', t_{n-1}) \\
             \cdot \pi(u', v', t_{n-1})\, \mathrm{d}{v'}.
        \end{multlined}
    \end{aligned}
    \label{eq:filt_prediction1}
\end{equation}

\subsubsection{The Filtering Distribution at Observation Points}
In this section, without loss of generality, we compute the filtering distribution at an observation time point $t_n$ as
\begin{equation}
    \begin{aligned}
        &\pi(u, v, t_n)\\
        &=\frac{\prob (u, v, t_n, \phi, y_{1:n})}{\prob (\phi, y_{1:n})} \\
        &=\frac{\prob (y_n \mid u, v, t_n, \phi, y_{1:n-1}) \prob (u, v, t_n, \phi, y_{1:n-1})} {\prob (y_n \mid \phi, y_{1:n-1}) \prob (\phi, y_{1:n-1})} \\
        &=Z_n^{-1}\prob (y_n \mid u, v)\pi(u, v, t_n^{-}),
    \end{aligned}
    \label{eq:filt_update_app}
\end{equation}
where $\pi(u, v, t_n^{-})=\prob (u, v, t_n\mid \phi, y_{1:n-1})$ is the filtering distribution at time $t_n$ before observation $y_n$ is added and
\begin{equation*}
    \begin{aligned}
        Z_n&= \prob (y_n \mid \phi, y_{1:n-1})\\
        &= \sum_{u \in \mathcal{U}} \int  \prob (y_n \mid u, v) \pi(u, v, t_n^{-})\, \mathrm  dv .
    \end{aligned}
\end{equation*}
Note that, \cref{eq:filt_update_app} is known as the update step of the filtering distribution.

\subsection{Calculation of the Backward Distribution}
\label{sec:backward}
We define the backward distribution as
\begin{equation*}
    \begin{aligned}
        \bt \coloneqq \prob (y_{n+1:K}\mid u, v, t, \phi),
    \end{aligned}
\end{equation*}
with probability measure
\begin{equation*}
    \begin{aligned}
        &\begin{multlined}
             \prob(y_{n+1:K}\mid u, v, t, \phi)\, \mathrm d y_{n+1:\nData}
        \end{multlined}\\
        &\begin{multlined}
             =\Prob(Y_{n+1:\nData} \in \mathrm d y_{n+1:\nData} \mid U(t)=u, \\
             V(t) \leq v, \Phi=\phi),
        \end{multlined}
    \end{aligned}
\end{equation*}
where $n=\max\{n' \in \mathbb{N}\mid t_{n'}\leq t\}$.
The calculation of the backward distribution can be split into two cases
\begin{inlineitemize}
    \item between observation time points and
    \item at observation time points.
\end{inlineitemize}

\subsubsection{The Backward Distribution Between Observation Points}
In this section, we compute the backward distribution in the time interval $[t-h, t]$ in which there is no observation
\begin{equation*}
    \begin{aligned}
        & \begin{multlined}
              \bh
        \end{multlined} \\
        & \begin{multlined}
              = \ph
        \end{multlined}\\
        &\begin{multlined}
             = \sum_{u'\in \mathcal{U}} \int \pph  \\
             \cdot \ppph \, \mathrm d v'.
        \end{multlined}
    \end{aligned}
\end{equation*}
Since the observations $Y_{n+1:\nData}$ given $U(t)$ and $V(t)$ are conditionally independent of $U(t-h)$ and $V(t-h)$, \ie,
\begin{equation*}
    \begin{aligned}
        &\pph\\
        &{}=\pppph
    \end{aligned}
\end{equation*}
we write
\begin{equation*}
    \begin{aligned}
        &\bh \\
        &=\sum_{u'\in \mathcal{U}} \int \bbh \ppph\, \mathrm d v'.
    \end{aligned}
\end{equation*}
This is the backward Chapman-Kolmogorov equation, see, \eg, \citet{kah:21},  for the probability distribution $ \prob (y_{n+1:K}\mid u, v, t, \phi)$.
Therefore, the backward distribution satisfies
\begin{equation*}
    \partial_t \bt = -\mathscr{A}^\dag \bt,
\end{equation*}
where the operator  $\mathscr{A}^\dag(\cdot )=\mathscr{D}^\dag (\cdot )+\mathscr{C}^\dag(\cdot)$ is given by
\begin{equation*}
    \begin{aligned}
        &\begin{multlined}
             \mathscr{D}^\dag \bt
             = \sum_{i \in \mathcal{D}} \kappa_{i}(u, v) (\bit  \\
             - \bt )
        \end{multlined}\\
        &\begin{multlined}
             \mathscr{C}^\dag \bt = \sum_{j \in \mathcal{C}} \kappa_{j}(u, v) \partial_{v_j}  \bt \\
             +\frac{1}{2} \sum_{j \in \mathcal{C}} \kappa_{j}(u, v) \partial_{v_j}^{2}  \bt.
        \end{multlined}
    \end{aligned}
\end{equation*}

\subsubsection{The Backward Distribution at Observation Points}
\label{sec:backs}

In this section, we calculate the backward distribution $\beta(u, v, t_{n+1}^{-})$ right before a observation point $t_{n+1}$ as follows
\begin{equation*}
    \begin{aligned}
        & \bthbb \\
        &=\pthn \\
        &=\frac{\pthnn}{\npth} \\
        &=\frac{\pthnnn}{\npth}\\
        &=\pthnnnn \\
        &\quad \cdot \frac{\nnpth}{\npth}\\
        &=\pthnnnn \\
        &\quad \cdot \pthnnnnn
    \end{aligned}
\end{equation*}

Letting $h \rightarrow 0$, we get
\begin{equation*}
    \begin{aligned}
        & \beta(u, v, t_{n+1}^{-})\\
        &= \lim_{h \rightarrow 0} \bthbb \\
        &= \beta(u, v, t_{n+1}) \prob (y_{n+1}\mid u, v, t_{n+1}, \phi).
    \end{aligned}
\end{equation*}

\subsection{Calculation of the Smoothing Distribution}
\label{sec:smth}
Assume we have all observations $y_{1:\nData}$ and we want to obtain the smoothing density $\tilde{\pi}(u, v, t)\coloneqq \prob (u, v, t \mid \phi, y_{1:\nData})$, that can be expressed as
\begin{equation*}
    \begin{aligned}
        & \tilde{\pi}(u, v, t)\\
        &=\frac{ \prob  (u, v, t, y_{1:n}, y_{n+1:\nData}, \phi)}{ \prob  (y_{1:n}, y_{n+1:\nData}, \phi)} \\
        &=\frac{ \prob  (y_{n+1:\nData}\mid u, v, t, \phi, y_{1:n} )}{ \prob  (y_{n+1:\nData} \mid \phi, y_{1:n})} \prob  ( u, v, t\mid \phi, y_{1:n})\\
        &=\frac{ \prob  (y_{n+1:\nData}\mid u, v, t, \phi )}{ \prob  (y_{n+1:\nData} \mid \phi, y_{1:n})} \prob  ( u, v, t \mid \phi, y_{1:n})\\
        &=\tilde Z_{n}^{-1} \beta (u, v, t) \pi (u, v, t).
    \end{aligned}
\end{equation*}
Note that the normalization constant
\begin{equation*}
    \begin{aligned}
        \tilde{Z}_n & \coloneqq \prob  (y_{n+1:\nData} \mid \phi, y_{1:n})\\
        &= \sum_{u \in \mathcal{U}} \int \beta(u, v, t) \pi(u, v, t) \, \mathrm dv,
    \end{aligned}
\end{equation*}
is almost surely constant~\citep{pardoux1981non}.
Next, we obtain the time derivative of the smoothing distribution.
We write
\begin{equation*}
    \begin{aligned}
        &\partial_t \smt \\
        &= \tilde{Z}_n^{-1} \partial_t \filt \bt\\
        &+ \tilde{Z}_n^{-1} \filt \partial_t \bt \\
        &= \tilde{Z}_n^{-1} \left[ \sum_{i \in \mathcal{D}}\left(\kiti \filti \right. \right. \\
        & \left. \left. -\kit \filt\right) \bt\right] \\
        &- \tilde{Z}_n^{-1} \sum_{j \in \mathcal{C}} \partial_{v_j} (\kt \filt) \bt \\
        &+ \tilde{Z}_n^{-1} \frac{1}{2} \sum_{j \in \mathcal{C}} \partial_{v_j}^{2} (\kt \filt ) \bt \\
        &+ \tilde{Z}_n^{-1} \sum_{i \in \mathcal{D}} \left[ \kit \left(\bt
        \right. \right. \\
        &\left. \left. -\bit \right) \filt \right] \\
        &- \tilde{Z}_n^{-1} \sum_{j \in \mathcal{C}} \kt \partial_{v_j}(\bt) \filt \\
        &- \tilde{Z}_n^{-1} \frac{1}{2} \sum_{j \in \mathcal{C}} \kt \partial^2_{v_j}(\bt)\filt
    \end{aligned}
\end{equation*}
\begin{equation*}
    \begin{aligned}
        &=\tilde{Z}_n^{-1} \sum_{i \in \mathcal{D}} \kiti \filti \bt \\
        &- \tilde{Z}_n^{-1} \sum_{j \in \mathcal{C}} \partial_{v_j} (\kt \filt ) \bt \\
        &+ \tilde{Z}_n^{-1} \frac{1}{2} \sum_{j \in \mathcal{C}} \partial_{v_j}^{2} (\kt \filt ) \bt \\
        &- \tilde{Z}_n^{-1} \sum_{i \in \mathcal{D}} \kit \bit \filt \\
        &- \tilde{Z}_n^{-1} \sum_{j \in \mathcal{C}} \kt \partial_{v_j}(\bt) \filt \\
        &- \tilde{Z}_n^{-1} \frac{1}{2} \sum_{j \in \mathcal{C}} \kt \partial^2_{v_j}(\bt) \filt. \\
    \end{aligned}
\end{equation*}
\newpage
By using
\begin{equation*}
    \tilde{Z}_n^{-1} \filt = \frac{\smt}{\bt}
\end{equation*}
we obtain
\begin{equation*}
    \begin{aligned}
        &\partial_t \smt \\
        &= \tilde{Z}_n^{-1} \sum_{i \in \mathcal{D}} \kiti \filti \bt \\
        &- \tilde{Z}_n^{-1} \sum_{j \in \mathcal{C}} \partial_{v_j} ( \kt \filt) \bt \\
        &+\tilde{Z}_n^{-1} \frac{1}{2} \sum_{j \in \mathcal{C}} \partial_{v_j}^2 (\kt \filt) \bt \\
        &- \sum_{i \in \mathcal{D}} \kit \frac{\bit}{\bt} \smt \\
        &- \sum_{j \in \mathcal{C}} \kt \partial_{v_j}(\bt) \frac{\smt}{\bt}\\
        &- \frac{1}{2} \sum_{j \in \mathcal{C}} \kt \partial^2_{v_j}(\bt) \frac{\smt}{\bt} \\
        &= \sum_{i \in \mathcal{D}} \kiti \smti \frac{\bt}{\bti}. \\
        &- \sum_{i \in \mathcal{D}} \kit \smt \frac{\bit}{\bt} \\
        &- \sum_{j \in \mathcal{C}} \partial_{v_j}( \kt \frac{\smt}{\bt}) \bt\\
        &+ \frac{1}{2} \sum_{j \in \mathcal{C}} \partial_{v_j}^{2}( \kt \frac{\smt}{\bt}) \bt \\
        &-\sum_{j \in \mathcal{C}} \kt \partial_{v_j}(\bt) \frac{\smt}{\bt} \\
        &- \frac{1}{2} \sum_{j \in \mathcal{C}} \kt \partial^2_{v_j} (\bt ) \frac{\smt}{\bt}. \\
    \end{aligned}
\end{equation*}

Now, we expand the derivatives as follows
    {\allowdisplaybreaks
\begin{align*}
    &\partial_{v_j} \left( \kt \frac{\smt}{\bt}\right) \\
    &= \bbt \partial_{v_j}(\kt) \smt \\
    &+\bbt \kt \partial_{v_j} (\smt) \\
    &- \btb \kt \smt \partial_{v_j}(\bt)\\
    &\partial_{v_j}^{2} \left( \kt \frac{\smt}{\bt}\right)\\
    &= \partial_{v_j} ( \partial_{v_j} (\kt \frac{\smt}{\bt} ) ) \\
    &= \partial_{v_j} \left[\partial_{v_j}(\kt) \smt \bbt \right]\\
    &+ \partial_{v_j} \left[\partial_{v_j}(\smt) \kt \bbt \right]\\
    &-\partial_{v_j} \left[ \partial_{v_j}(\bt) \kt \smt \btb \right]\\
    &= \partial^{2}_{v_j}(\kt) \frac{\smt}{\bt} \\
    &+ \partial_{v_j}(\smt) \partial_{v_j}(\kt) \bbt\\
    &+\partial^{2}_{v_j}(\smt) \kt \bbt \\
    &+\partial_{v_j}(\smt) \partial_{v_j}(\kt) \bbt \\
    &- \btb \\
    &\cdot \left[\partial_{v_j}(\bt) \partial_{v_j}(\smt ) \kt \right. \\
    &+ \partial_{v_j}(\bt) \partial_{v_j}(\kt ) \smt \\
    &+ \partial_{v_j}^{2}(\bt) \kt \smt \\
    &+ \partial_{v_j}(\bt) \partial_{v_j}(\kt) \smt \\
    &+\left. \partial_{v_j}(\bt) \kt \partial_{v_j} (\smt) \right] \\
    &+ 2 \bbbt \left( \partial_{v_j}(\bt) \partial_{v_j}(\bt) \right. \\
    & \left. \kt \smt \right).
\end{align*}}

Then, we get
    {\allowdisplaybreaks
\begin{align*}
    &\partial_{t} (\smt)  \\
    &=\sum_{i \in \mathcal{D}} \kiti \smti \frac{\bt}{\bti} \\
    &-\sum_{i \in \mathcal{D}} \kit \smt \frac{\bit}{\bt} \\
    &-\sum_{j \in \mathcal{C}} \partial_{v_j}(\kt)  \smt +\kt  \\
    &\qquad\cdot \partial_{v_j}(\smt) \\
    &+ \sum_{j \in \mathcal{C}} \kt \smt \partial_{v_j}(\bt)\\
    &\qquad \cdot \bbt \\
    &+ \frac{1}{2} \sum_{j \in \mathcal{C}}
    \left[ \partial^{2}_{v_j}(\kt) \smt \right. \\
    &+ \partial_{v_j}(\smt) \partial_{v_j}(\kt) + \partial^{2}_{v_j}(\smt)      \\
    &\qquad \cdot \kt\\
    &+ \left. \partial_{v_j}(\smt) \partial_{v_j}(\kt) \right] \\
    &- \frac{\bbt}{2} \sum_{j \in \mathcal{C}} \left[ \partial_{v_j}(\bt) \right.  \\
    &\qquad \cdot\partial_{v_j}(\smt) \kt \\
    &+ \partial_{v_j}(\bt) \partial_{v_j}(\kt) \smt \\
    &+ \partial_{v_j}^{2}(\bt) \kt \smt \\
    &+ \partial_{v_j}(\bt) \partial_{v_j}(\kt) \smt \\
    &+ \left. \partial_{v_j}(\bt) \kt \partial_{v_j}(\smt) \right] \\
    &+ \btb \sum_{j \in \mathcal{C}} \left[ \partial_{v_j}(\bt) \partial_{v_j}(\bt) \right. \\
    &\quad \left. \kt \smt \right] \\
    &-\sum_{j \in \mathcal{C}}\kt \partial_{v_j}(\bt) \frac{\smt}{\bt}\\
    &- \frac{1}{2} \sum_{j \in \mathcal{C}} \kt \partial^2_{v_j} (\bt ) \frac{\smt}{\bt} \\
    &=- \sum_{j \in \mathcal{C}} \partial_{v_j}(\kt \smt ) \\
    &+ \sum_{j \in \mathcal{C}} \partial_{v_j}^2(\kt \smt) \\
    &- \frac{\bbt}{2}\sum_{j \in \mathcal{C}} \left[2 \partial_{v_j}(\bt)\right. \\
    &\cdot \partial_{v_j}(\smt) \kt \\
    &+ 2 \partial_{v_j}(\bt) \partial_{v_j}(\kt) \smt \\
    &+\left. 2 \partial_{v_j}^2(\bt) \kt \smt \right] \\
    &+ \btb \sum_{j \in \mathcal{C}} \partial_{v_j}^2(\bt) \smt \kt \\
    &+ \sum_{i \in \mathcal{D}} \kiti \smti \frac{\bt}{\bti} \\
    &- \sum_{i \in \mathcal{D}} \kit \smt \frac{\bit}{\bt}.
\end{align*}}

By using the product rule, we write
    {\allowdisplaybreaks
\begin{align*}
    &\partial_{v_j}( \frac{\partial_{v_j}(\bt) \smt \kt}{\bt})\\
    &=\left( \partial^2_{v_j}(\bt) \smt \kt \right. \\
    &+\partial_{v_j}(\bt) \partial_{v_j}(\smt) \kt \\
    &+\left. \partial_{v_j}(\bt) \partial_{v_j}(\kt) \smt \right) \\
    &\qquad \cdot\bbt \\
    &- (\partial_{v_j}(\bt))^2 \smt \kt \btb.
\end{align*}}
This gives us
    {\allowdisplaybreaks
\begin{align*}
    &\partial_{t} (\smt) \\
    &= -\sum_{j \in \mathcal{C}} \partial_{v_j}(\kt \smt )\\
    & + \sum_{j \in \mathcal{C}}
    \partial_{v_j}^2(\kt \smt) \\
    &-\sum_{j \in \mathcal{C}} \partial_{v_j}( \frac{\partial_{v_j}(\bt) \smt \kt}{\bt}) \\
    &+ \sum_{i \in \mathcal{D}} \kiti \smti \frac{ \bt}{\bti} \\
    &- \sum_{i \in \mathcal{D}} \kit \smt \frac{\bit}{\bt} \\
    &=- \sum_{j \in \mathcal{C}} \partial_{v_j}\left( \kt \smt \right. \\
    &+\left. \frac{\partial_{v_j}(\bt) \smt \kt}{\bt}\right) \\
    &+ \sum_{j \in \mathcal{C}} \partial_{v_j}^2 (\kt \smt ) \\
    &+ \sum_{i \in \mathcal{D}} \kiti \smti \frac{ \bt}{\bti} \\
    &- \sum_{i \in \mathcal{D}} \kit \smt \frac{\bit}{\bt} \\
    &=- \sum_{j \in \mathcal{C}} \partial_{v_j} \left(\{ \kt +\partial_{v_j} \log (\bt)   \right. \\
    &\left. \quad \quad \quad \cdot \kt \} \smt \right) \\
    &+ \sum_{j \in \mathcal{C}} \partial_{v_j}^2 (\kt \smt ) \\
    &+ \sum_{i \in \mathcal{D}} \kiti \smti \frac{ \bt}{\bti} \\
    &- \sum_{i \in \mathcal{D}} \kit \smt \frac{ \bit}{\bt}. \\
\end{align*}}

\section{State Inference}
\label{sec:state_inference_appendix}

\subsection{Forward-Filtering and the Bootstrap Filter}
\label{sec:bootstrap_appendix}
Consider that we want to sample from the following posterior distribution
\begin{equation}
    \begin{multlined}
        U_{[0, t_n]}, V_{[0, t_n]} \mid Y_{1:n}, \Phi\\
        \sim \Prob (\mathrm d u_{[0, t_n]}, \mathrm d v_{[0, t_n]} \mid y_{1:n}, \phi).
    \end{multlined}    \label{eq:posterior_measure_importance_der}
\end{equation}
By exploiting the model structure from
\cref{sec:model}
this posterior distribution can be expressed as
\begin{equation*}
    \begin{aligned}
        & \Prob (\mathrm d u_{[0, t_n]}, \mathrm d v_{[0, t_n]} \mid y_{1:n}, \phi) \\
        &\propto \prob (y_n \mid u_{[0, t_n]}, v_{[0, t_n]}, y_{n-1}) \\
        &\cdot \Prob ( \mathrm d u_{[0, t_n]}, \mathrm d v_{[0, t_n]} \mid y_{1:n-1}, \phi)\\
        &{} =\prob (y_n \mid u_n, v_n)\\
        &\cdot \Prob ( \mathrm d u_{[t_{n-1}, t_n]}, \mathrm d v_{[t_{n-1}, t_n]} \mid u_{n-1}, v_{n-1}, \phi)\\
        &\cdot \Prob (\mathrm d u_{[0, t_{n-1}]}, \mathrm d v_{[0, t_{n-1}]}\mid y_{1:n-1}, \phi).
    \end{aligned}
\end{equation*}
Next, we want to sample from this distribution using importance sampling.
By using a proposal distribution $\mathrm{Q}(\mathrm d u_{[0, t_n]}, \mathrm d v_{[0, t_n]}\mid y_{1:n}, \phi)$ we produce $M$ particles
\begin{equation*}
    \begin{multlined}
        U_{[0, t_n]}^{(i)}, V_{[0, t_n]}^{(i)} \mid Y_{1:n}, \Phi \\
        \sim \mathrm{Q}( \mathrm d u_{[0, t_n]}^{(i)}, \mathrm d v_{[0, t_n]} ^{(i)}\mid y_{1:n}, \phi),
    \end{multlined}
\end{equation*}
with $i=1, 2, \dots, M.$

The corresponding weight $\Gamma_n^{(i)}$ of the $i$th particle is then given by
\begin{equation}
    \begin{aligned}
        \Gamma_n^{(i)} &\propto \prob (y_n \mid u_n^{(i)}, v_n^{(i)})\\ &\frac{ \Prob (\mathrm d u_{[t_{n-1}, t_n]}^{(i)}, \mathrm d v_{[t_{n-1}, t_n]}^{(i)} \mid u_{n-1}^{(i)}, v_{n-1}^{(i)}, \phi)}{\mathrm{Q}(\mathrm d u_{[0, t_n]}^{(i)}, \mathrm d v_{[0, t_n]}^{(i)}\mid y_{1:n}, \phi)}\\
        & \cdot \Prob (\mathrm d u_{[0, t_{n-1}]}^{(i)}, \mathrm d v_{[0, t_{n-1}]}^{(i)}\mid y_{1:n-1}, \phi)
    \end{aligned}
    \label{eq:importance_weight_filter_der}
\end{equation}

For a proposal factorizing as
\begin{equation*}
    \begin{split}
        &\mathrm{Q}(\mathrm d u_{[0, t_n]}^{(i)}, \mathrm d v_{[0, t_n]}^{(i)}\mid y_{1:n}, \phi)\\
        &{}=\mathrm{Q}( \mathrm d u_{[t_{n-1}, t_n]}^{(i)}, \mathrm d v_{[t_{n-1}, t_n]}^{(i)} \mid u_{n-1}^{(i)}, v_{n-1}^{(i)}, y_{1:n}, \phi)
        \\  &\quad  \cdot
        \mathrm{Q}(\mathrm d u_{[0, t_{n-1}]}^{(i)}, \mathrm d v_{[0, t_{n-1}]}^{(i)}\mid y_{1:n-1}, \phi),
    \end{split}
\end{equation*}
\cref{eq:importance_weight_filter_der} can be written recursively as
\begin{equation*}
    \begin{aligned}
        &{\Gamma}_{n}^{(i)} \propto \prob (y_n \mid u_n^{(i)}, v_n^{(i)})
        \\ &\cdot \frac{\Prob (\mathrm d u_{[t_{n-1}, t_n]}^{(i)}, \mathrm d v_{[t_{n-1}, t_n]}^{(i)} \mid u_{n-1}^{(i)}, v_{n-1}^{(i)}, \phi)}{\mathrm{Q}(\mathrm d u_{[t_{n-1}, t_n]}^{(i)}, \mathrm d v_{[t_{n-1}, t_n]}^{(i)}\mid u_{n-1}^{(i)}, v_{n-1}^{(i)}, y_{1:n}, \phi)}\\
        & \cdot {\Gamma}_{n-1}^{(i)}
    \end{aligned}
\end{equation*}

If we now choose the proposal distribution to be the dynamics of the prior evolution, \ie,
\begin{equation*}
    \begin{aligned}
        &\mathrm{Q}( \mathrm d u_{[t_{n-1}, t_n]}^{(i)}, \mathrm d v_{[t_{n-1}, t_n]}^{(i)} \mid u_{n-1}^{(i)}, v_{n-1}^{(i)}, y_{1:n}, \phi)\\
        &{}=\Prob (\mathrm d u_{[t_{n-1}, t_n]}^{(i)}, \mathrm d v_{[t_{n-1}, t_n]}^{(i)} \mid u_{n-1}^{(i)}, v_{n-1}^{(i)}, \phi),
    \end{aligned}
\end{equation*}
we end up with the \emph{bootstrap filter} \citep{dfg:01, Gordon_1993}, for which the weights can be easily computed as
\begin{equation*}
    \Gamma_{n}^{(i)} \propto \prob (y_n \mid u_n^{(i)}, v_n^{(i)}) \Gamma_{n-1}^{(i)}.
\end{equation*}

Given this particle description, the filtering distribution at time point $t_n$ is hence approximated as
\begin{equation*}
    \prob (u, v, t_n \mid y_{1:n})\approx \sum_{i=1}^M \Gamma_n^{(i)} \delta(U^{(i)}_n-u)\delta(V^{(i)}_n-v).
\end{equation*}
The bootstrap filter computes the weights recursively, by sampling from the particle distribution.
It uses the prior distribution as the proposal distribution and it replaces particles having low-importance weights with other particles having high-importance weights.
This method is practical, as it can be easily implemented for many complex systems.
The method is based on three steps which are \emph{initialization}, \emph{importance resampling}, and \emph{selection}.
In the rest of this section, we explain the details of these steps.

\subsubsection*{First Step: Initialization.}
In the presented model, at iteration step $n=0$, the process $(U,V)$ starts at $t=t_0=0$, with the particles $U_0^{(i)}=V_0^{(i)}=0$, and equal weights ${\Gamma}_{0}^{(i)}=M^{-1}$, for all particles $i=1, 2, \dots, M$.
This yields a particle-based version of the initial condition using the empirical measure as
\begin{equation*}
    \prob (u, v, t_0) = \sum_{i=1}^M {\Gamma}_{0}^{(i)} \delta(U^{(i)}_0-u)\delta(V^{(i)}_0-v),
\end{equation*}
with particles $\{(U_0^{(i)}, V_0^{(i)})\}_{i=1, \dots, M}$ and weights ${\Gamma}_{0}=( {\Gamma}_{0}^{(1)}, {\Gamma}_{0}^{(2)}, \dots, {\Gamma}_{0}^{(M)} )^T$.

Next, for the iteration steps $n=1, 2, \dots, \nData$ we perform the importance sampling step and the selection step.

\subsubsection*{Second Step: Importance Sampling.}
For all particles $i=1, 2, \dots, M$, we sample from the prior dynamics, \ie,
\begin{equation*}
    \begin{aligned}
        &(U_{[t_{n-1}, t_n]}^{(i)}, V_{[t_{n-1}, t_n]}^{(i)}) \\
        &\sim \Prob ( \mathrm d u_{[t_{n-1}, t_n]}^{(i)}, \mathrm d v_{[t_{n-1}, t_n]}^{(i)} \mid u_{n-1}^{(i)}, v_{n-1}^{(i)}, \phi)
    \end{aligned}
\end{equation*}

Using this set of particles an approximation to the distribution $\prob (u, v, t_{n} \mid y_{1:n-1}, \phi)$ can be build as
\begin{equation*}
    \begin{aligned}
        &\prob (u, v, t_{n} \mid y_{1:n-1}, \phi) \\
        &{}\approx \sum_{i=1}^M \frac{1}{M} \delta(U^{(i)}_n-u)\delta(V^{(i)}_n-v),
    \end{aligned}
\end{equation*}
similar to the prediction step in \cref{eq:filt_prediction1}.

Next, we compute the to unity normalized weights ${\Gamma}_n=({\Gamma}_{n}^{(1)}, {\Gamma}_{n}^{(2)}, \dots, {\Gamma}_{n}^{(M)})$ as
\begin{equation*}
{\Gamma}
    _{n}^{(i)} \propto \prob (y_n\mid {u}_n^{(i)}, {v}_n^{(i)}).
\end{equation*}
These weights give an approximation for the posterior distribution $\prob (u, v, t_n \mid y_{1:n}, \phi)$ as
\begin{equation*}
    \prob (u, v, t_n \mid y_{1:n})\approx \sum_{i=1}^M \Gamma_n^{(i)} \delta(U^{(i)}_n-u)\delta(V^{(i)}_n-v),
\end{equation*}
similar to the update step in \cref{eq:filt_update_app}.

\subsubsection*{Third Step: Selection.}
To avoid degeneracy which can be seen very often in filtering algorithms, we compute the \emph{effective sample size}
\begin{equation}
    \label{eq:ess}
    \mathrm{ESS}=\left( \sum_{i=1}^{M} ({\Gamma}_{n}^{(i)})^2 \right)^{-1}.
\end{equation}

If $\mathrm{ESS} \leq \alpha M$ where $0< \alpha \leq 1$ is a user-defined constant specifying the minimum effective particle ratio, see, \eg, \citet{MIHAYLOVA20141}, \citet{dj:11}, \citet{liu2008}, and \citet{spe:16}, we resample the filtered particles $\{U_{[t_{n-1}, t_n]}^{(i)}, V_{[t_{n-1}, t_n]}^{(i)}\}_{i=1}^M$.
In this resampling phase, based on an appropriate resampling algorithm,
we replicate the particles with a high weight ${\Gamma}_{n}^{(i)}$, while particles with lower weights are eliminated.
This gives a particle-based approximation for the posterior distribution $\prob (u, v, t_n \mid y_{1:n}, \phi)$, with equal weights $\Gamma_{n}^{(i)}=M^{-1}$.
There are three widely used resampling algorithms, which are \emph{systematic resampling}, \emph{residual resampling}, and \emph{multinomial resampling}.
In this work, we use systematic resampling.

As a summary of the forward-filtering step, we update the given particles recursively in the forward direction by using the system equation.
Then, we resample the particles using weights proportional to the observation likelihood to generate filtered particles.
In the following section, we explain the details of how to obtain the smoothed particles by using the filtered particles in this step.

\subsection{Backward Smoothing}
\label{sec:backward smoothing appendix}
Next, consider that we want to generate samples from the posterior distribution
\begin{equation*}
    \begin{multlined}
        U_{[0, T]}, V_{[0, T]} \mid Y_{1:K}, \Phi\\
        \sim \Prob(\mathrm d u_{[0, T]}, \mathrm d v_{[0, T]} \mid y_{1:K}, \phi)
    \end{multlined}
\end{equation*}
by using the particle trajectories $\{U_{[0, T]}^{(i)}, V_{[0, T]}^{(i)}\}_{i=1}^M$ obtained from the bootstrap filter. The particles are importance samples distributed according to the posterior path measures in \cref{eq:posterior_measure_importance_der}. The weights of the particles at the last time step can be hence calculated similarly to \cref{eq:importance_weight_filter_der} as
\begin{equation*}
    \begin{aligned}
        &\tilde{\Gamma}^{(i)} \propto \Prob ( \mathrm d u_{[t_{K}, T]}^{(i)}, \mathrm d v_{[t_{K}, T]}^{(i)}\mid u_{K}^{(i)}, v_{K}^{(i)}, \phi) \\
        &\cdot \prob (y_K \mid u_K^{(i)}, v_K^{(i)})\\
        &\cdot \frac{\Prob ( \mathrm d u_{[t_{K-1}, t_K]}^{(i)}, \mathrm d v_{[t_{K-1}, t_K]}^{(i)} \mid u_{K-1}^{(i)}, v_{K-1}^{(i)}, \phi)}{\mathrm{Q}( \mathrm d u_{[0, T]}^{(i)}, \mathrm d v_{[0, T]}^{(i)}\mid y_{1:K}, \phi)}\\
        &\quad \cdot \Prob ( \mathrm d u_{[0, t_{K-1}]}^{(i)}, \mathrm d v_{[0, t_{K-1}]}^{(i)}\mid y_{1:K-1}, \phi)\\
        &=\frac{\Prob ( \mathrm d u_{[t_{K}, T]}^{(i)}, \mathrm d v_{[t_{K}, T]}^{(i)}\mid u_{K}^{(i)}, v_{K}^{(i)}, \phi)}{\mathrm{Q}(\mathrm d u_{[t_{K}, T]}^{(i)}, \mathrm d v_{[t_{K}, T]}^{(i)}\mid u_{K}^{(i)}, v_{K}^{(i)}, y_{1:K}, \phi)} \\
        & \cdot \prob (y_K \mid u_K^{(i)}, v_K^{(i)})\\
        &\cdot \frac{\Prob ( \mathrm d u_{[t_{K-1}, t_K]}^{(i)}, \mathrm d v_{[t_{K-1}, t_K]}^{(i)} \mid u_{K-1}^{(i)}, v_{K-1}^{(i)}, \phi)}{\mathrm{Q}( \mathrm d u_{[t_{K-1}, t_{K}]}^{(i)}, \mathrm d v_{[t_{K-1}, t_{K}]}^{(i)}\mid u_{K-1}^{(i)}, v_{K-1}^{(i)}, y_{1:K}, \phi)}\\
        & \cdot \Gamma_{K-1}^{(i)}
    \end{aligned}
\end{equation*}

As we choose the importance distribution as
\begin{equation*}
    \begin{aligned}
        &{\mathrm{Q}(\mathrm d u_{[t_{K}, T]}^{(i)}, \mathrm d v_{[t_{K}, T]}^{(i)}\mid u_{K}^{(i)}, v_{K}^{(i)}, y_{1:K}, \phi)} \\
        &={\Prob (\mathrm d u_{[t_{K}, T]}^{(i)}, \mathrm d v_{[t_{K}, T]}^{(i)}\mid u_{{K-1}}^{(i)}, v_{{K-1}}^{(i)}, \phi)} \\
        &\mathrm{Q}(\mathrm d u_{[t_{K-1}, t_{K}]}^{(i)}, \mathrm d v_{[t_{K-1}, t_{K}]}^{(i)}\mid u_{K-1}^{(i)}, v_{K-1}^{(i)}, y_{1:K}, \phi)\\
        &=\Prob ( \mathrm d u_{[t_{K-1}, t_K]}^{(i)}, \mathrm d v_{[t_{K-1}, t_K]}^{(i)} \mid u_K^{(i)}, v_K^{(i)}, \phi),
    \end{aligned}
\end{equation*}
we have that the smoothing weight can be computed as
\begin{equation}
    \begin{aligned}
        \tilde{\Gamma}^{(i)}\propto \prob (y_K\mid u_K^{(i)}, u_K^{(i)}) \Gamma_{K-1}^{(i)}
        \propto {\Gamma}^{(i)}_K.
    \end{aligned}
    \label{eq:smoothing_weight_sis}
\end{equation}

Hence, a sample of the desired posterior distribution can be evaluated by sampling from the particle approximation
\begin{equation*}
    \begin{aligned}
        &\Prob ( \mathrm d u_{[0, T]}, \mathrm d v_{[0, T]} \mid \phi, y_{1:K}) \\&\approx \sum_{i=1}^M {\Gamma}^{(i)}_K \delta_{U_{[0, T]}^{(i)}}(\mathrm d u_{[0, T]})\delta_{V_{[0, T]}^{(i)}}(\mathrm d v_{[0, T]}),
    \end{aligned}
\end{equation*}
with weights $\Gamma_K=( {\Gamma}^{(1)}_K, {\Gamma}^{(2)}_K, \dots, {\Gamma}^{(M)}_K)^\top$, \ie,

\begin{equation*}
    \begin{split}
        &U_{[0, T]}, V_{[0, T]} \mid \Phi, Y_{1:K} \\
        &\sim \sum_{i=1}^M {\Gamma}^{(i)}_K \delta_{U_{[0, T]}^{(i)}}(\mathrm d u_{[0, T]})\delta_{V_{[0, T]}^{(i)}}(\mathrm d v_{[0, T]}).
    \end{split}
\end{equation*}
This strategy is known as \ac{sir} particle smoothing \citep{kit:96, sarkka_2013}.

\section{Experiments}
\label{sec:app_experiments}
The joint density function $\prob (u, v, t \mid \mathcal{H})$ representing the time-point wise marginal distribution of \cref{eq:count_v} satisfies the following \ac{hme}
{\allowdisplaybreaks
    \begin{align*}
        &\partial_t \prob (u, v, t \mid \mathcal{H})\\
        &{}=- \partial_v ( \kappa_1 (u, v) \prob (u, v, t \mid \mathcal{H}) ) \\
        &\quad+\frac{1}{2} \partial_v^2 ( \kappa_1 (u, v) \prob (u, v, t \mid \mathcal{H}) )\\
        &\quad+\kappa_2(u-1, v) \prob (u-1, v, t \mid \mathcal{H})\\
        &\quad -\kappa_2(u, v) \prob (u, v, t \mid \mathcal{H})\\
        &{}= -\partial_v ( \kappa_1 (u, v) ) \prob (u, v, t \mid \mathcal{H})\\
        &\quad- \kappa_1 (u, v) \partial_v ( \prob (u, v, t \mid \mathcal{H})) \\
        &\quad+ \partial_v( \prob (u, v, t \mid \mathcal{H}) )\partial_v(\kappa_1 (u, v))\\
        &\quad+\frac{1}{2} \partial_v^2 ( \kappa_1 (u, v) )\prob (u, v, t \mid \mathcal{H}) \\
        &\quad+\frac{1}{2} \partial_v^2 ( \prob (u, v, t \mid \mathcal{H}) )\kappa_1 (u, v)\\
        &\quad+\kappa_2(u-1, v) \prob (u-1, v, t \mid \mathcal{H})\\
        &\quad-\kappa_2(u, v) \prob (u, v, t \mid \mathcal{H}) \\
        &{}=\kappa_1 (u, v)(\frac{1}{2} \partial_v^2 ( \prob (u, v, t \mid \mathcal{H}) )-\partial_v \prob (u, v, t \mid \mathcal{H}) )\\
        &\quad + \partial_v(\kappa_1 (u, v)) (\partial_v \prob (u, v, t \mid \mathcal{H}) -\prob (u, v, t \mid \mathcal{H}) ) \\
        &\quad+\frac{1}{2} \partial_v^2 ( \kappa_1 (u, v) )\prob (u, v, t \mid \mathcal{H})\\
        &\quad+\kappa_2(u-1, v) \prob (u-1, v, t \mid \mathcal{H})\\
        &\quad-\kappa_2(u, v) \prob (u, v, t \mid \mathcal{H}).
    \end{align*}}
    \end{appendices}
   
    \bibliography{hybrid_bibliography}

\end{document}